\documentclass[a4paper,11pt]{article}

\usepackage{amssymb}
\usepackage{jinstpub}
\usepackage{graphicx}

\usepackage{mhchem}

\usepackage{lineno}
\usepackage{float}
\usepackage{longtable}
\usepackage[hang]{subfigure}	
\usepackage{caption}

\usepackage{array}
\newcolumntype{H}{>{\setbox0=\hbox\bgroup}c<{\egroup}@{}}

\begin{document}

\title{Comparative characterization study of LYSO:Ce crystals for timing applications}

\author[a,b,1]{F. M. Addesa,\note{Corresponding author.}}
\author[b]{P.Barria,}
\author[b]{R. Bianco,}
\author[b]{M. Campana,}
\author[b]{F. Cavallari,}
\author[c]{A. Cemmi,}
\author[d]{M. Cipriani,}
\author[b]{I. Dafinei,}
\author[b]{B. D'Orsi,}
\author[b]{D. del Re,}
\author[b]{M. Diemoz,}
\author[b]{G. D'Imperio,}
\author[b]{E. Di Marco,}
\author[c]{I. Di Sarcina,}
\author[e]{M. Enculescu,}
\author[b]{E. Longo,}
\author[f]{M. T. Lucchini,}
\author[g]{F. Marchegiani,}
\author[b]{P. Meridiani,}
\author[g]{S. Nisi,}
\author[b]{G. Organtini,}
\author[b]{F. Pandolfi,}
\author[b]{R. Paramatti,}
\author[b]{V. Pettinacci,}
\author[b]{C. Quaranta,}
\author[b]{S. Rahatlou,}
\author[b]{C. Rovelli,}
\author[b]{F. Santanastasio,}
\author[b]{L. Soffi,}
\author[b]{R. Tramontano,}
\author[a]{C. G. Tully}
\emailAdd{faddesa@princeton.edu, Francesca.Addesa@roma1.infn.it}

\affiliation[a]{Princeton University, Princeton, U.S.A.}
\affiliation[b]{INFN sezione di Roma and Sapienza University, Roma, Italy}
\affiliation[c]{ENEA\,-\,Fusion and Technology for Nuclear Safety and Security Department, Rome, Italy}
\affiliation[d]{CERN, European Organization for Nuclear Research, Geneva 23, CH-1211 Switzerland }
\affiliation[e]{National Institute of Materials Physics, Magurele, Romania}
\affiliation[f]{INFN and University of Milano-Bicocca, Milano, Italy}
\affiliation[g]{INFN - Laboratori Nazionali del Gran Sasso,  Assergi L’Aquila, Italy}


\arxivnumber{2205.14890 } 
\abstract{
Cerium-doped Lutetium-Yttrium Oxyorthosilicate (LYSO:Ce)
is one of the most widely used Cerium-doped Lutetium based scintillation crystals. Initially developed for medical detectors it rapidly became attractive for High Energy Particle Physics (HEP) applications, especially in the frame of high luminosity particle colliders. 

In this paper, a comprehensive and systematic study of LYSO:Ce
($[Lu_{(1-x)}Y_x]_2SiO_5$:$Ce$) crystals is presented. It involves for the first time a large number of crystal samples (180) of the same size from a dozen of producers. 
The study consists of a comparative characterization of LYSO:Ce crystal products available on the market by mechanical, optical and scintillation measurements and aims specifically, to investigate key parameters of timing applications for HEP.}


\maketitle
\flushbottom
\section{Introduction}
\label{S:1}

Cerium-doped Lutetium-Yttrium Oxyorthosilicate, commonly known as LYSO:Ce, is one of the most widely used Cerium-doped Lutetium based scintillation crystals.
Initially developed for medical applications \cite{patent,1239590}, in particular for Positron Emission Tomography (PET), its characteristics in terms of high mass density (twice the density of NaI(Tl)), fast scintillation kinetics (6 times faster decay time than BGO) and high light yield (40000\,ph/MeV) attracted also the interest of the High Energy Physics (HEP) community. 

In the last decade, LYSO:Ce was employed to prototype and realize high precision electromagnetic calorimeters such as the one designed for the Mu2e experiment \cite{Mu2e:2012eea} and the CCALT forward calorimeter of the KLOE-2 experiment \cite{Cordelli_2011}. 

More recently, a new crystal R\&D effort driven by the requirement for high time resolution of second generation PET (Time of Flight PET) further improved the performance of LYSO leading to the industrial production of faster crystals (decay time $<40$\,ns) and with higher light yield than in the past \cite{da7b17043ca547b49732c9654549d667, 04e31a8d8be64fa497420d7a3528f3e3}. The latter, together with the excellent resistance to $\gamma$ radiation \cite{7429799}, neutrons \cite{5402125} and charged hadrons \cite{8305635}, makes LYSO appealing for timing applications in the harsh environment of future high-luminosity particle colliders.  
Here, the high rate of simultaneous interactions per bunch crossing (\textit{pileup}) will produce spatial overlap of tracks and energy deposits. This will affect the capability to disentangle physics events through the traditional detector layers. A picosecond timing layer dedicated to time of arrival measurement of charged particles can help to associate tracks to the correct vertex, mitigating the pileup effect.  
In this context, the CMS experiment at the Large Hadron collider (LHC) chose LYSO:Ce crystals coupled to Silicon Photomultipliers (SiPMs) to design the sensor unit for the barrel part (BTL) of its timing layer, the MIP Timing Detector \cite{CMS:2667167}. With this layout, the BTL will be able to provide precision timing of minimum ionizing particles with a resolution of 30\,-\,60\,ps \cite{Abbott_2021} restoring the event reconstruction performance of the pre high luminosity era.

In this paper, a comparative and systematic study of LYSO:Ce crystal properties is carried out, for the first time for a wide number of crystal samples and crystal manufacturers and with particular attention to the key features responsible for the timing performance of the crystals. The aim is to offer a comprehensive review of the state of the art of LYSO:Ce crystal products currently available on the market and identify the best producers for the BTL project.
The performance of LYSO:Ce crystals have been evaluated using bare crystal samples (without wrapping) studying the following properties:
\begin {itemize}
\item mass density and correlation with Yttrium content;
\item optical transmission characteristics and evaluation of the $Ce^{3+}$ relative concentration;
\item photoluminescence characteristics;
\item light output and decay time;
\item light yield and decay time temperature dependency (low temperature range);
\item $\gamma$ radiation resistance.
\end {itemize}
\section{Samples}
\label{S:2}
LYSO:Ce crystal samples from 12 manufacturers were studied and compared. A list of the manufacturers is provided below in alphabetic order. Each one is randomly associated with a number from 1 to 12 which identifies the producer's crystals throughout this work. Therefore the id number does not match with the order in the following list. 
\begin {itemize}
\item Crystal Photonics, USA
\item EPIC Crystal, China
\item Hamamatsu Photonics, Japan 
\item Hypercrystal\,-\,NSYSU, Taiwan
\item Saint-Gobain, France 
\item Shanghai EBO Optoelectronics, China 
\item Shanghai Institute of Ceramics, China
\item Simcrystals Technology, China
\item SIPAT, China
\item Suzhou JT Crystal Technology, China
\item Tianle Photonics, China
\item Zecotek Imaging System, Singapore
\end {itemize}
The LYSO:Ce ($[Lu_{(1-x)}Y_x]_2SiO_5$:$Ce$) crystals analyzed have a variable Yttrium and Cerium content depending on the manufacturer. Both are related to fundamental properties of the crystals. The Yttrium content correlates with the mass density and consequently with the MIP deposited energy, while the Cerium content is related to light yield and decay time. Dedicated measurements were performed to determine Yttrium and $Ce^{3+}$ concentrations and are described in the following paragraphs.
\subsection{Sample description}
\label{SS:2-1}
\begin{figure}[ht]
\begin{center}
\includegraphics [width = 0.45 \textwidth] {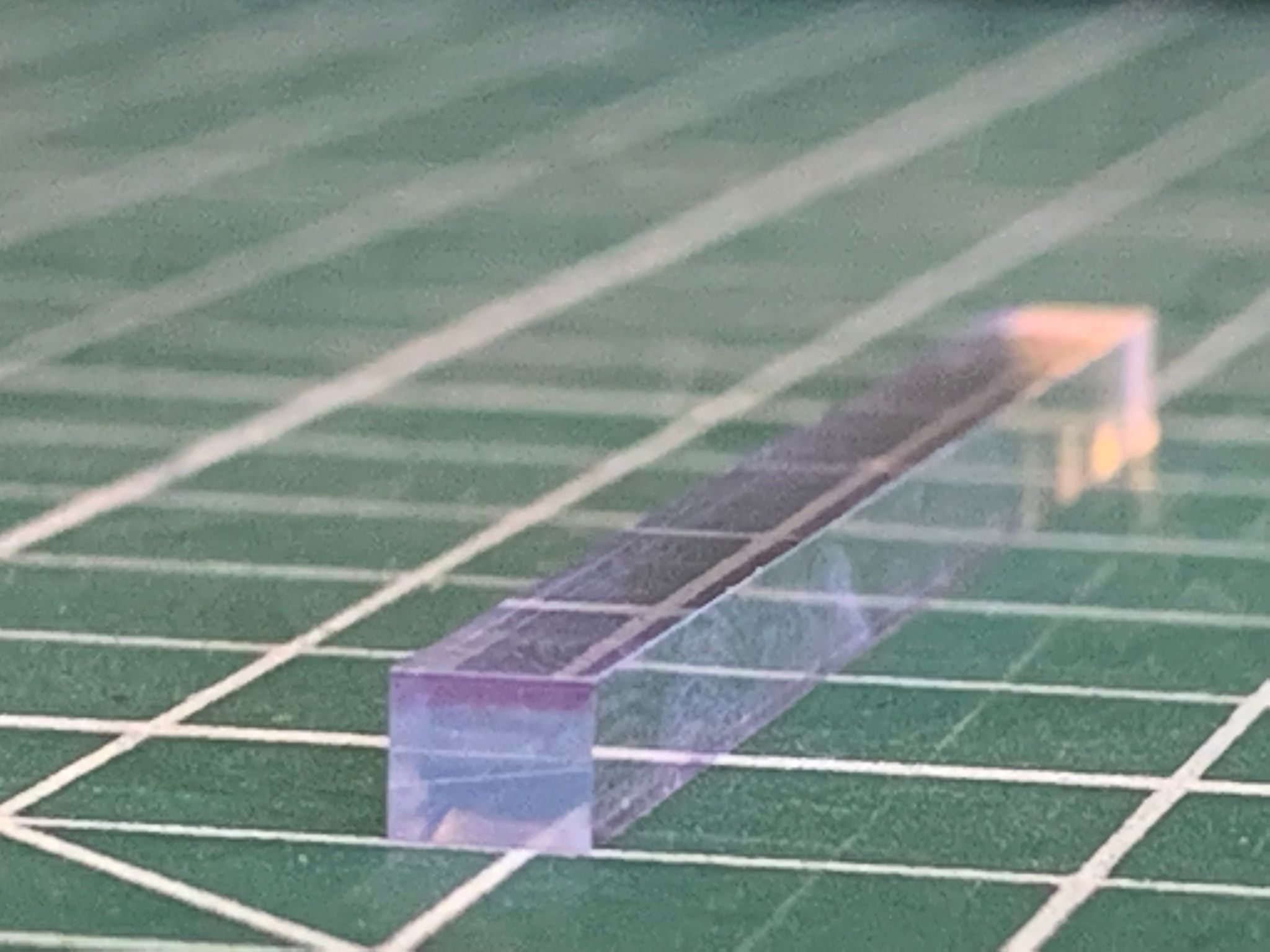}
\includegraphics [width = 0.45 \textwidth] {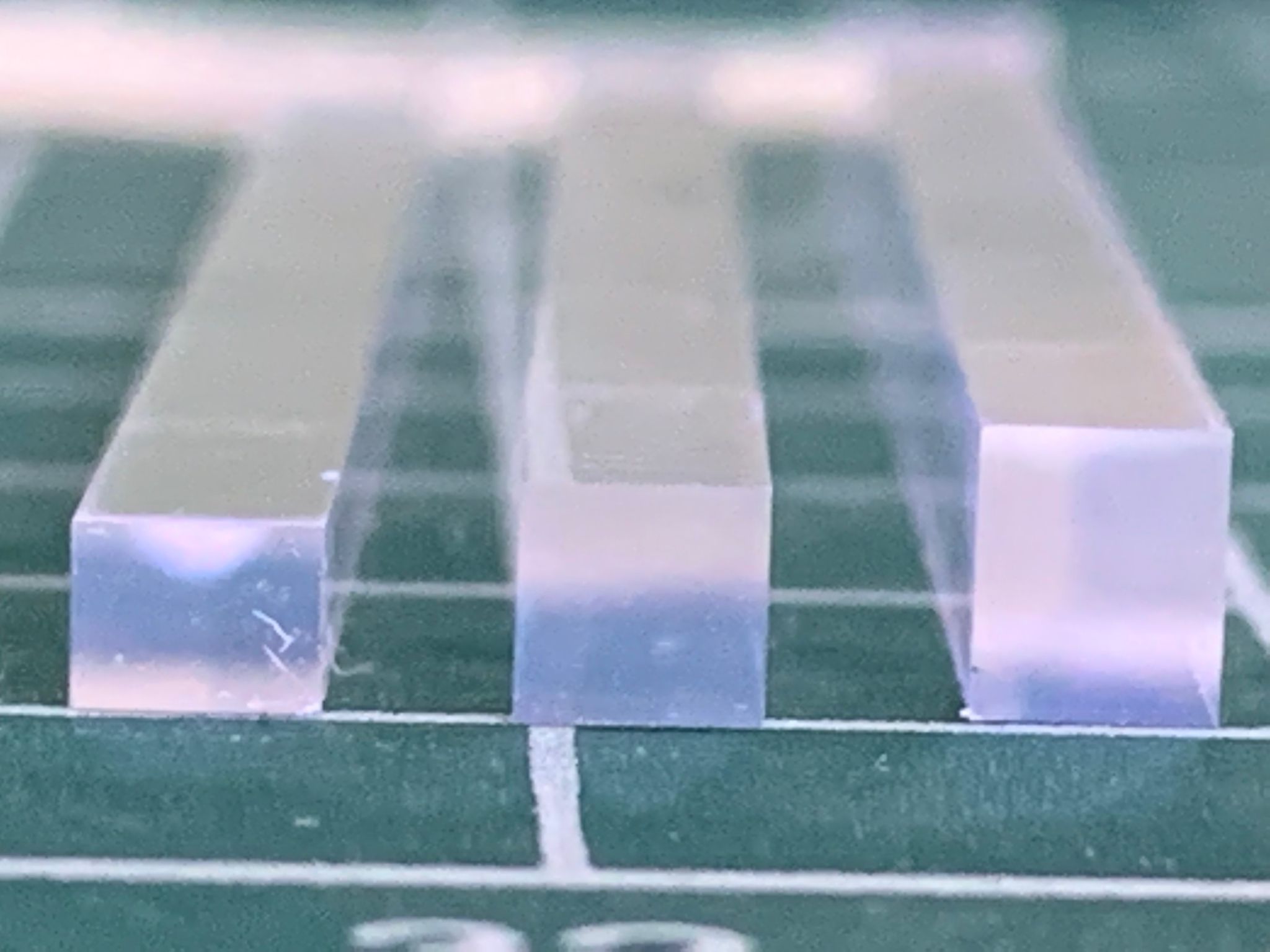}
\caption{Example of a crystal bar sample (left). Cross-sectional view of the 3 types of bars used in this study (right).}
\label{fig:LYSO}
\end{center}
\end{figure}

The LYSO:Ce samples studied are 57\,mm long crystal bars. The section is rectangular with 3 different thicknesses. The nominal dimensions are reported in Table \ref{tab:bar_parameters} for the 3 geometries.
All manufacturers provided 15 crystal bars, 5 for each geometry and all cut from the same ingot. Samples were provided with an optical surface quality of $Ra<$15\,nm for all six faces. 
Most of the crystal properties were measured for all the samples of a manufacturer.
When the set of crystals analyzed is smaller or with different characteristics, it is reported in detail.
Fig.~\ref{fig:LYSO} shows an example of a crystal bar sample (left) and a cross-sectional view of the 3 different available geometries (right).

\begin{table}[!ht]
\begin{center}
\begin{tabular}{|c|c|c|c|c|c|c|}
\hline
{\bf geometry} & \multicolumn{3}{|c|}{\bf bar dimension (mm)}  & {\bf \# of samples}\\
\cline{2-4}
{\bf type} & {\bf w} & {\bf t} & {\bf L} & {\bf per producer}\\
\hline
{\bf 1} & 3.12  & 3.75 & 57.00 & 5 \\
{\bf 2} & 3.12 & 3.00 & 57.00 & 5\\
{\bf 3} & 3.12 & 2.40 & 57.00  & 5 \\
\hline
\end{tabular}
\caption{Nominal dimensions of the crystal bars. The bar width, thickness and length are labeled respectively as w, t, and L.  }
\label{tab:bar_parameters}
\end{center}
\end{table}

\subsection{Density of LYSO:Ce samples}
 \label{SS:2.2}
 The density of the LYSO:Ce bars is calculated based on the measurements of dimensions and mass. The crystal density is an indicator of the Yttrium percentage in the crystal composition, as shown later in this section, and is directly related to the amount of deposited energy by a Minimum Ionizing Particle (MIP) crossing the crystal.
 
 \subsubsection*{Dimensions measurement}
  \label{SSS:2.2.1}
\begin{figure}[htbp!!!]
\begin{center}
\subfigure[\label{mitutoyo}]{\includegraphics [width = 0.24 \textwidth, height=0.28 \textwidth] {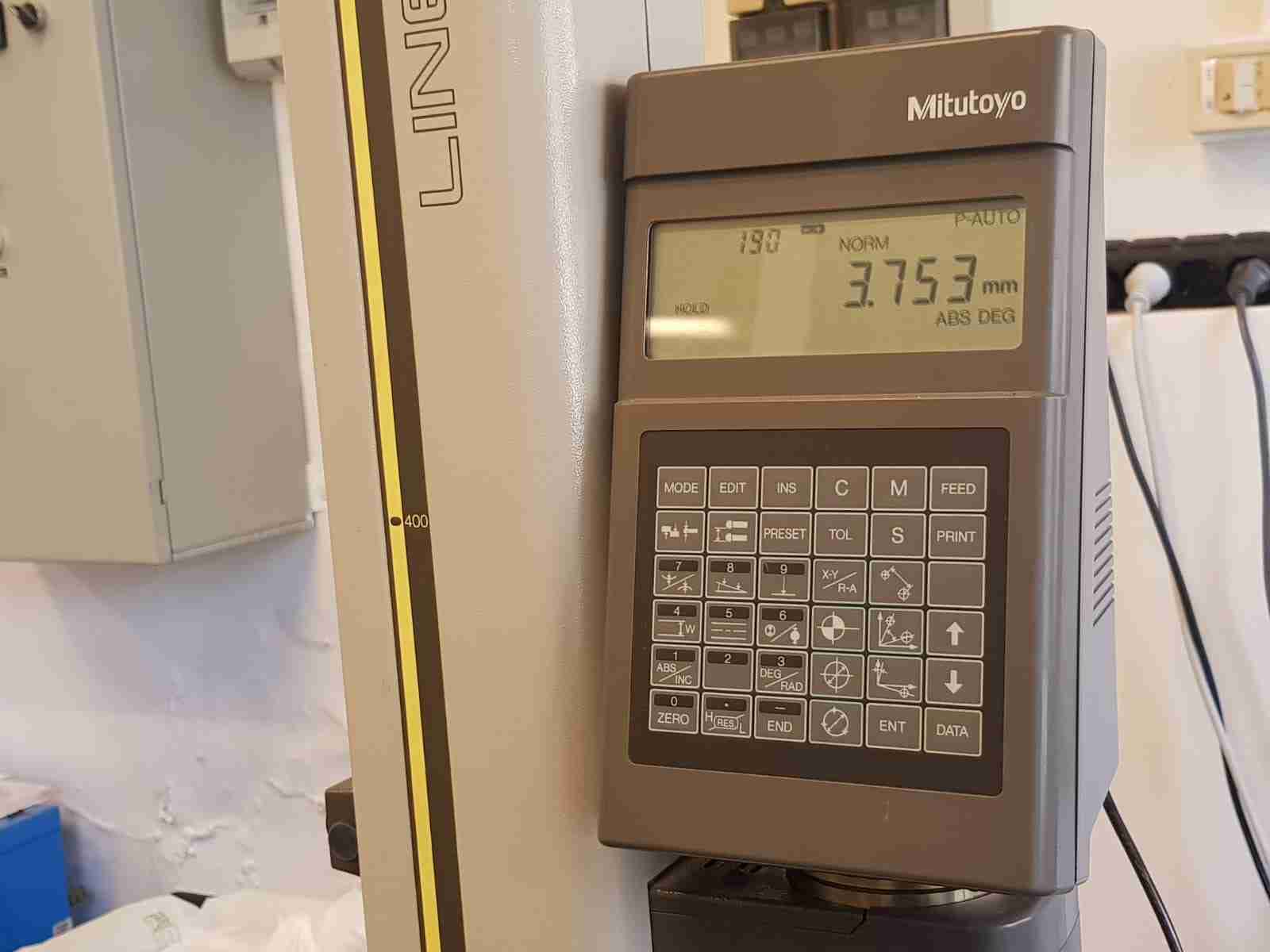}}
\subfigure[\label{mit_wt}]{\includegraphics [width = 0.24 \textwidth, height=0.28 \textwidth] {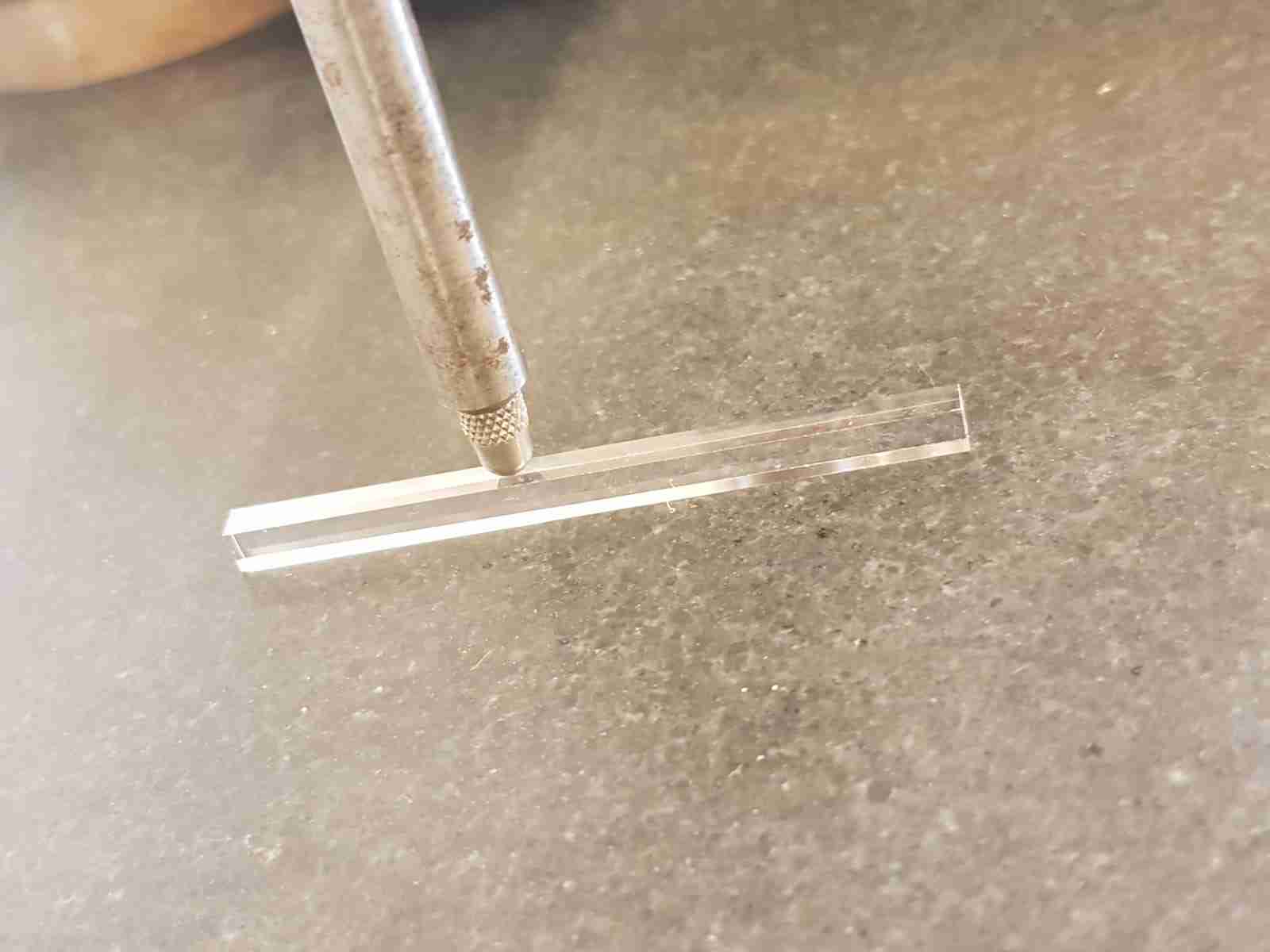}}
\subfigure[\label{mit_lenght}]{\includegraphics [width = 0.24 \textwidth, height=0.28 \textwidth] {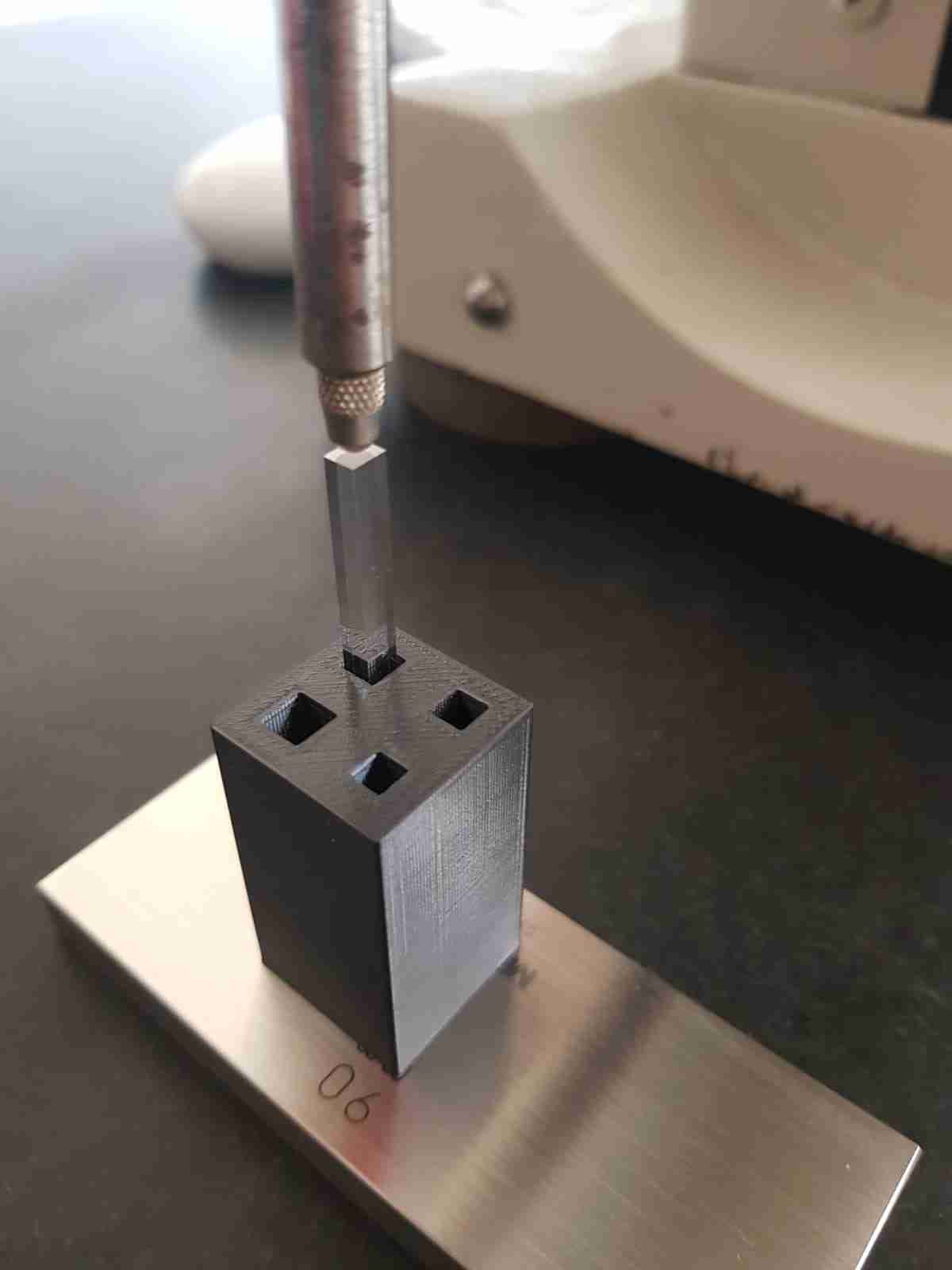}}
\subfigure[\label{balance}]{\includegraphics [width = 0.24 \textwidth, height=0.28 \textwidth] {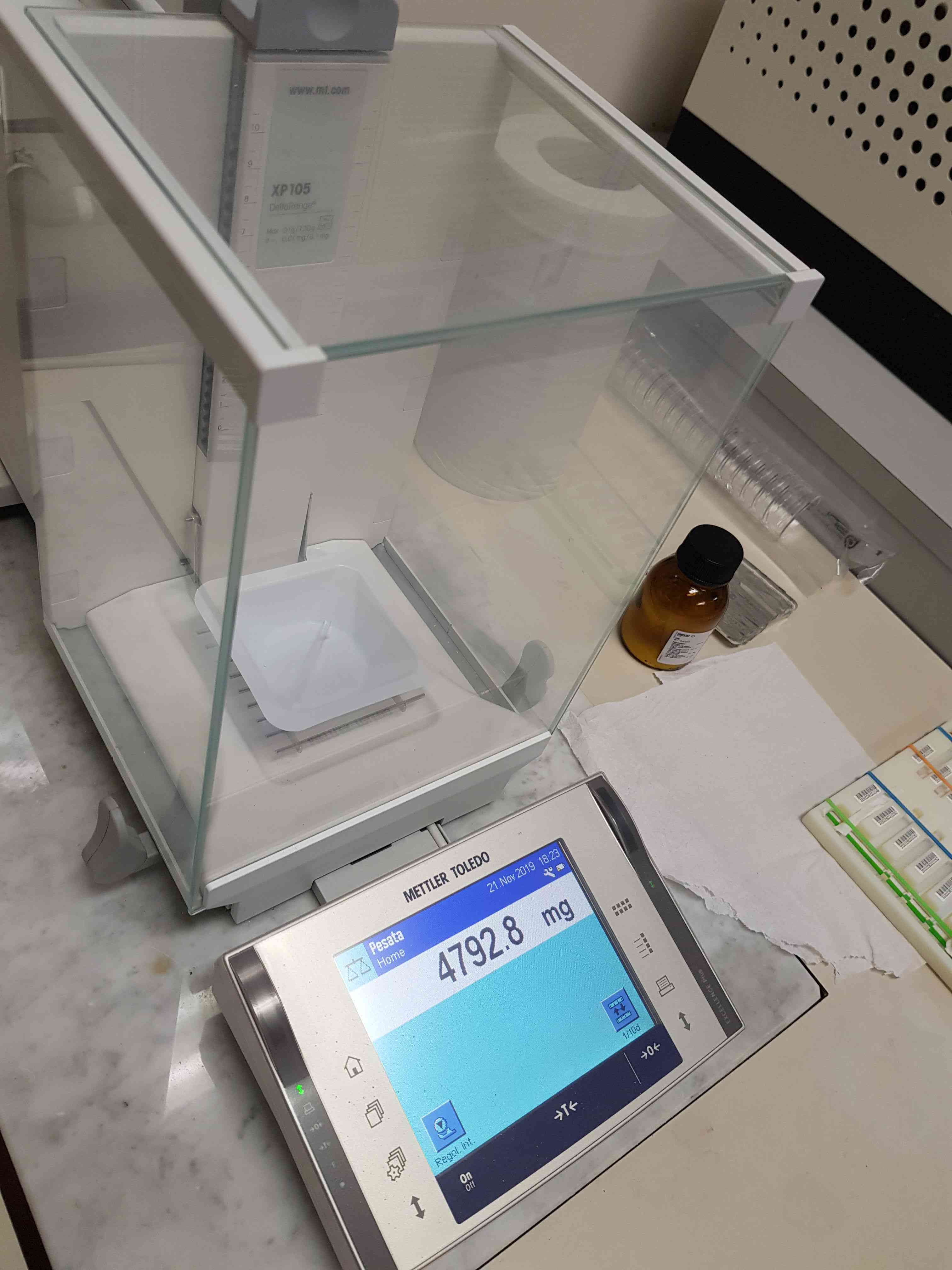}}
\caption{ Mitutoyo LH-600 measurement system to measure the crystal bar dimensions (a); measurement of $w$ and $t$(b) and $L$ dimension of the crystal bar with the Mitutoyo probe (c); Mettler Toledo XP105 balance for the mass measurements of the crystal bars (d).}
\end{center}
\end{figure}

   A high performance measurement system Mitutoyo LH-600 (Fig.~\ref{mitutoyo}) was used to measure the three dimensions of the crystal bars. The digital resolution of the instrument is 1\,$\mu$m and the observed reproducibility of the measurement is 2-3\,$\mu$m. The measurements were carried out on a flat granite table in a temperature controlled environment at T$\sim 20^\circ$C ($\pm 1 ^\circ$C).

\begin{figure}[!htp!]
\begin{center}
\includegraphics[width = 0.49 \textwidth, height=0.40 \textwidth]{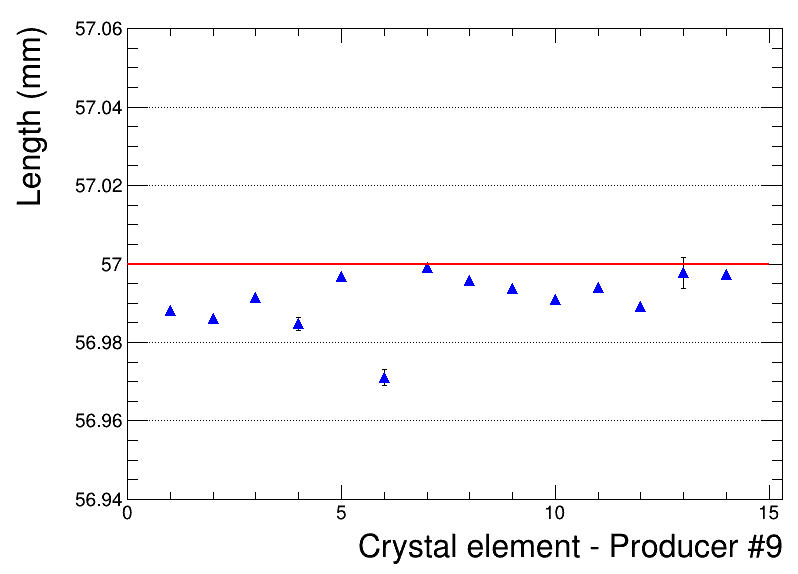}
\includegraphics[width = 0.49 \textwidth, height=0.40 \textwidth]{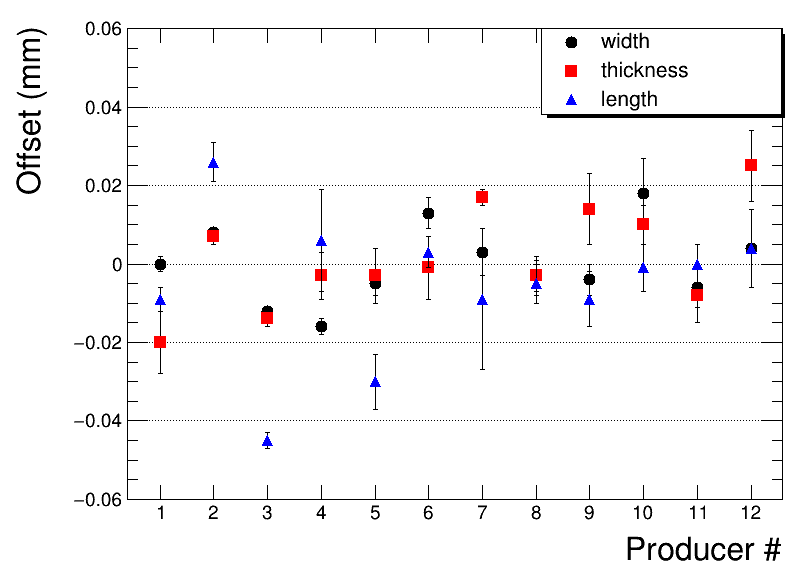}
\caption{Measured $L$ for all crystal samples of producer 9. The red line corresponds to the nominal crystal length (left). Measured offsets with respect to the nominal dimensions of Table \ref{tab:bar_parameters}. The mean and the standard deviation of bar dimension offsets are shown for each producer. L data point for producer 9 corresponds to the average of the L values shown in the left plot subtracted by the nominal L value (right).}
\label{fig:size}
\end{center}
\end{figure}

For each single bar, width ($w$) and thickness ($t$) are defined as the average of 16 measurements in different positions along the crystal axis (Fig.~\ref{mit_wt}), while the length ($L$) as the average of 8 measurements made in the 4$\times$2 corners of the ends of the bars, as shown in Fig.~\ref{mit_lenght} . The black 3D printed holder, with holes of different transverse size, was used to support vertically the bar without any pressure on it and to avoid accidental falls. In Fig.~\ref{fig:size} (left), $L$ is shown for all the crystal elements of producer 9. The data points and the error bars correspond to the average and the standard deviation of the 8 measurements performed to determine L respectively.
With the purpose to add information related to the compliance with the dimension specifications of the samples provided by each producer, the results are given showing the offsets defined as the difference between the measured values and the nominal values reported in Table \ref{tab:bar_parameters}. Moreover, to characterize the uniformity of the samples, the offset values are averaged over the 15 crystals of each producer and the errors bars are the related standard deviations (Fig.~\ref{fig:size}, right).
All the producers show a good mastering of the cutting technology. For almost all of them, the standard deviation of each dimension is within 5\,$\mu$m while the tolerance with respect to the nominal dimensions is within 30\,$\mu$m.

\subsubsection*{Mass measurements}
\label{SSS:2.2.2}
The mass measurement of the crystal bars was performed with the high-precision Mettler Toledo XP105 balance (0.1 mg digital resolution). The reproducibility of the measurements is better than 0.5 mg; the balance is provided with a glass enclosure for protection against drafts (Fig.~\ref{balance}).
The measurements were carried out in a temperature controlled environment at T$\sim 20^\circ$C ($\pm 1 ^\circ$C). The mean values and the standard deviation of the measured bar masses are given in Table 2 for each crystal geometry and for all the producers.
\begin{table}[H]
\begin{center}
\begin{tabular}{|c|l|l|l|}
\hline
Producer & Type 1 & Type 2 & Type 3 \\
 & (mg) & (mg) & (mg) \\
\hline
1 & 4713\,$\pm$\,8 & 3749\,$\pm$\,11 & 2995\,$\pm$\,3 \\
2 & 4760\,$\pm$\,2 & 3805\,$\pm$\,1 & 3043\,$\pm$\,3 \\
3 & 4795\,$\pm$\,3 & 3833\,$\pm$\,11 & 3060\,$\pm$\,1 \\
4 & 4622\,$\pm$\,6 & 3730\,$\pm$\,3 & 2943\,$\pm$\,5 \\
5 & 4721\,$\pm$\,2 & 3778\,$\pm$\,4 & 3017\,$\pm$\,3 \\
6 & 4765\,$\pm$\,4 & 3800\,$\pm$\,13 & 3014\,$\pm$\,14 \\
7 & 4906\,$\pm$\,14 & 3921\,$\pm$\,9 & 3148\,$\pm$\,4 \\
8 & 4782\,$\pm$\,7 & 3816\,$\pm$\,13 & 3055\,$\pm$\,6 \\
9 & 4738\,$\pm$\,4 & 3777\,$\pm$\,15 & 3041\,$\pm$\,4 \\
10 & 4935\,$\pm$\,11 & 3938\,$\pm$\,11 & 3169\,$\pm$\,9 \\
11 & 4734\,$\pm$\,6 & 3771\,$\pm$\,7 & 3024\,$\pm$\,4 \\
12 & 4765\,$\pm$\,36 & 3839\,$\pm$\,8 & 3079\,$\pm$\,5 \\
\hline
\end{tabular}
\caption{LYSO bar mass per crystal geometry and producer. The mean and the standard deviation of the mass values are reported for each geometry.}
\label{tab:masses}
\end{center}
\end{table}
\subsubsection*{Density measurements}
\label{SSS:2.2.3}
The density value is calculated by dividing the mass of the bar by its volume as calculated from the measured dimensions. The density uncertainty is obtained by the corresponding uncertainties on dimensions and mass (the latter being negligible). Results are summarized in Fig.~\ref{fig:density} (left) where the density, between 7.0 and 7.4\,g/cm$^3$, is shown as the mean over the 15 crystals of the same producer. The error bar corresponds to the relative standard deviation (standard deviation over the mean) which is well below 1\,\% for all the producers.
\subsubsection*{Yttrium fraction with ICP-MS measurements and density correlation}
\label{SSS:2.2.4}
The chemical formula of the Cerium-doped LYSO crystals of this study is $[Lu_{(1-x)}Y_x]_2SiO_5$:$Ce$. The stoichiometry of  ($[Lu_{(1-x)}Y_x]$) group is not fixed and depends on the crystal growth recipe of each manufacturer (expected values for x are below 10\,\%).
The large difference between the atomic mass of Lutetium (174.967\,amu) and Yttrium (88.906\,amu) leads to significant differences between the densities of LYSO crystals having different Yttrium content. 
The Yttrium molar fraction for at least one crystal bar from each producer was measured by Inductively Coupled Plasma Mass Spectrometry (ICP-MS), at the Gran Sasso National Laboratory (LNGS, Aquila, Italy). For one of the producers, a set of 6 crystals were measured in order to check the consistency of the measurement within the same producer. In total, 31 crystal bars were measured by the ICP-MS technique.

\begin{figure}[!ht]
\begin{center}
\includegraphics[width = 0.49 \textwidth, height=0.40 \textwidth]{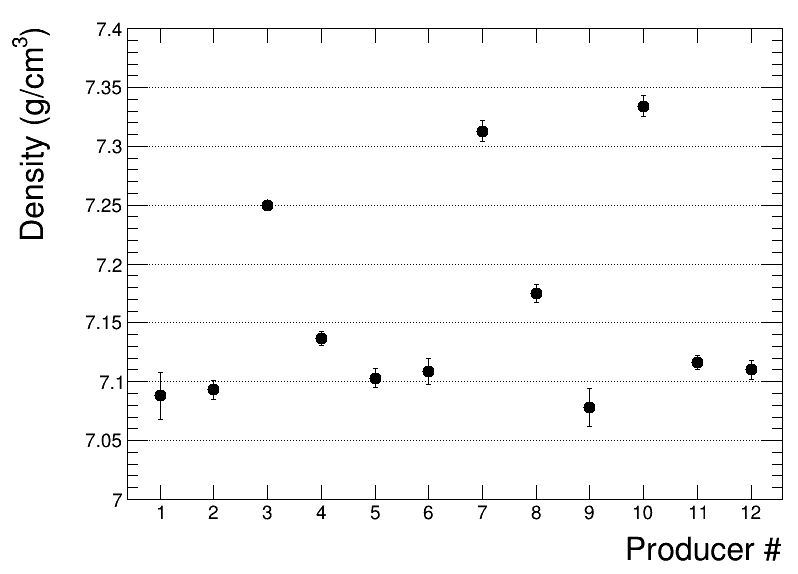}
\includegraphics[width = 0.49 \textwidth, height=0.4 \textwidth]{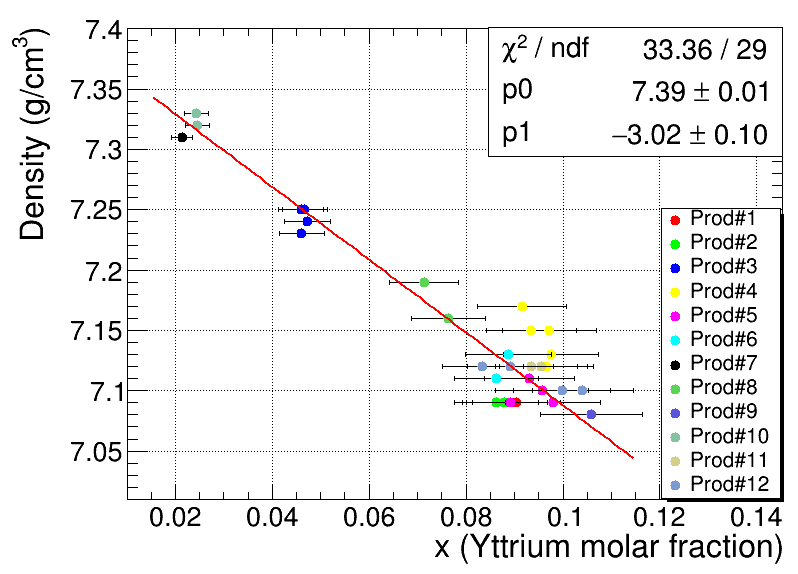}
\caption{The mean density of the 15 bars is shown for each producer. The relative standard deviation of the bar density is well below 1\,\% for all producers (left). Crystal density as a function of the Yttrium molar fraction ($x$) for a subsample of the crystal bars analyzed in this study. The linear correlation is clearly visible (right).}
\label{fig:density}
\end{center}
\end{figure}
The results showing the Yttrium content and its linear correlation with the measured mass density are reported in Fig.~\ref{fig:density} (right). Measurements from all the crystals of the subsample analyzed are shown and correspond to a data point. The linear correlation of the Yttrium fraction of a crystal bar with its density is clearly demonstrated and the linear regression coefficient is $R=0.95$. In addition, a linear fit with $\chi^2$ minimization has been applied to the data.
The linear fit parameters correspond, within the error, to the empirical linear relation of the Yttrium content and the density of the crystal which can be determined by the densities of pure LSO ($x=0$, density= $7.4\,g/cm^3$) and pure YSO ($x=1$, density = $4.5\,g/cm^3$) crystals.
\section{Optical properties}
\label{SS:3}
\subsection{Transmission}
\begin{figure}[!ht]
\begin{center}
\includegraphics[width = 0.70 \textwidth, height=0.45 \textwidth]{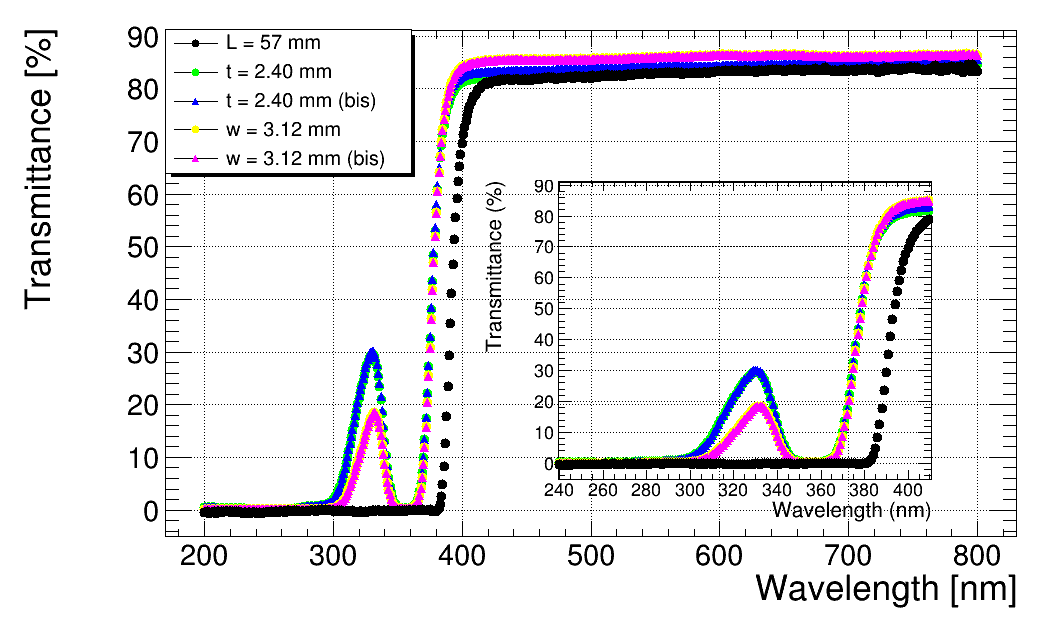}
\caption{Optical transmission spectra measured in all three directions: longitudinal (black dots), and transversal along $w$ (yellow dots and violet triangles) and $t$ (green dots and blue triangles). Two measurements for each transverse direction are displayed.}
\label{fig:Transm-new.JPG}
\end{center}
\end{figure}

Transmission spectra were measured along the three directions of the crystal samples, one longitudinal ($L$) and two transversal along width ($w$) and thickness ($t$). The measurements were performed at room temperature in the range 300\,-\,800\,nm. Fig.~\ref{fig:Transm-new.JPG} gives an example of transmission spectra measured for one crystal in all three directions. The figure shows on one side the reproducibility of the transmission measurement and on the other hand it illustrates the nature of the transmission threshold in the UV region. The transmission threshold is caused by the Cerium doping and not by the fundamental absorption of LYSO. Undoped LYSO crystals are indeed transparent in a wider range, with the fundamental absorption at 200\,nm at room temperature \cite{Ricci-2010}.\\ The transverse dimensions of the samples (2\,-\,3\,mm) did not allow for the study of the region below 300\,nm due to saturating absorption on color centers induced by dopants (mainly Ce). Double beam spectrophotometers were used: P.E. Lambda 950 at CERN and UV–Vis–NIR CARY 5000 (Varian, Agilent Technologies Deutschland GmbH) at NIMP Bucharest.

Although the sample dimensions were unsuitable (too thick) for a detailed analysis of the optical absorption, the spectra measured in the transverse directions ($w$ and $t$) allowed for the visualization of the $5d^1$ absorption band of $Ce^{3+}$ and implicitly the evaluation of the relative Cerium concentration in the measured crystals. To this purpose, the absorbance spectra in the region of interest (ROI) from 440\,nm down to 300\,nm (2.8-4.1\,eV) were obtained from the transmission spectra (Fig.~\ref{fig:Transm-new.JPG}, zoom) and fitted using a function which takes into consideration the main absorption centers acting in that ROI.
In the considered ROI, the absorbance is found to be proportional to the absorption coefficient ($\alpha$) and the sample transverse size ($d$):
\begin{equation} \label{eq3}
A\sim \alpha \cdot d
\end{equation}

\noindent Details about how Eq.~\ref{eq3} was analytically obtained by the transmission expression are provided in \ref{A}.

The absorption coefficient can be decomposed into the sum of the contributions from different absorption centers $j$, each one described by an absorption coefficient $\alpha_{j}$, which is proportional to the concentration $N_{j}$ of the respective absorption center:

\begin{equation} \label{eq4}
\begin{split}
\alpha & = \sum_{j}\alpha_{j} \\
 & where:    \alpha _{j}=k_{j}\cdot N_{j} \\
\end{split}
\end{equation}

In the defined ROI, $\alpha$ can be written as:

\begin{equation} \label{eq6}
\alpha=\alpha_{Ce^{3+}}+\alpha_{other}
\end{equation}
 
\noindent where $\alpha_{Ce^{3+}}$ represents the contribution of $Ce^{3+}$ absorption centers while $\alpha_{other}$ takes into account the contribution of all the other absorption centers.

\begin{figure}[!htp!]
\begin{center}
\includegraphics[width = 0.495 \textwidth, height=0.40 \textwidth]{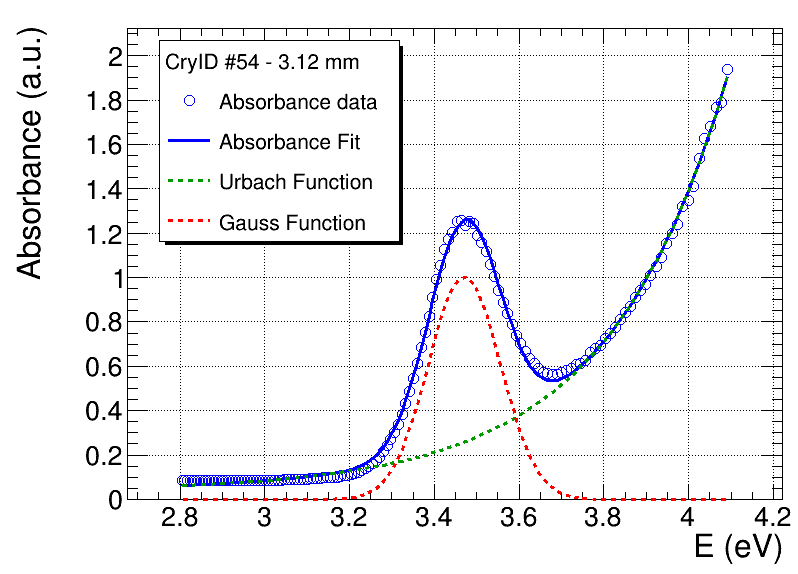}
\includegraphics[width = 0.495 \textwidth, height=0.40 \textwidth]{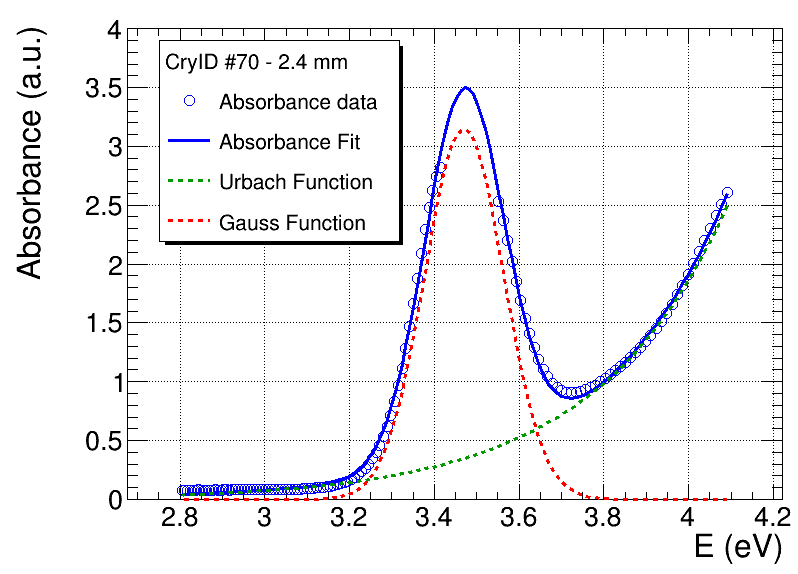}
\caption{Absorbance spectra for different crystals with the applied fit (blue line). The contributes to the fitting function coming from $Ce^{3+}$( red dotted line) and all the other (green dotted line) absorbing centers are also shown.  
}
\label{fig:FitAbsorbance}
\end{center}
\end{figure}

The parameter $\alpha_{Ce^{3+}}$ is described by a Gaussian function of the energy 
The amplitude of the Gaussian function is proportional to the concentration of Cerium in the sample.
The absorption due to all the other absorbing centers ($\alpha_{other}$) can be described by an empirical exponential function, similar to that applied in the Urbach approximation \cite{{Urbach-1953}, {Keil-2010}}:


 The ratio between the amplitude of the Gaussian function and the sample width can be used for a relative estimation of the concentration of $Ce^{3+}$ centers in the sample ($N_{Ce^{3+}}$).  The fit function is effective for all the spectra, regardless of the Cerium doping and possible co-doping used by different crystal producers, as illustrated in Fig.~\ref{fig:FitAbsorbance}. 

Transmission spectra were measured for 39 crystals from different producers with at least two crystals from each producer. For producer 4, 5 and 6, samples from different ingots and with different declared Cerium concentration were studied. The corresponding $N_{Ce^{3+}}$ value are reported in Tab.~\ref{tab:ce_content}. A total of 23 crystals were measured in both transversal directions, $w$ and $t$, and often more than one measurement was taken for a given direction, thus having a total of 75 optical transmission spectra analyzed. This was made in order to check both the reproducibility of the transmission spectrum measurement and the overall stability of the ($N_{Ce^{3+}}$) measurement procedure.

\begin{table}
\begin{minipage}[t]{0.495\linewidth}
\begin{tabular}{|c|c|l|l|}
\hline
Prod. & Sample\,\# & $N_{Ce^{3+}}$ & Lab.\\
\hline
1 &	1 & 1.7540 & CERN\\
1 &	2 & 1.4990 & CERN\\
1 &	3 & 1.2230	& NIMP\\
1 &	4 & 1.2450	& NIMP\\
2 &	1 & 2.1010 & CERN\\
2 &	2 & 1.4590	& CERN\\
2 &	3 & 1.5520	& NIMP\\
3 &	1 & 0.3244	& CERN\\
3 &	2 & 0.3231	& CERN\\
3 &	3 & 0.3240	& NIMP\\
4 &	1 & 1.9800	& CERN\\
4 &	2 & 1.2480	& NIMP\\
4 &	3 & 1.5990	& CERN\\
4 &	4 & 0.6741 & NIMP\\
5 &	1 & 0.3481	& CERN\\
5 &	2 & 0.2560 & NIMP\\
5 &	3	& 0.3779 & CERN\\
5 &	4 & 0.4304 & NIMP\\
6 &	1 & 1.2850 & CERN\\
6 &	2 & 1.1040 & CERN\\
\hline
\end{tabular}
\end{minipage}
\hspace{0.001\linewidth}
\begin{minipage}[t]{0.495\linewidth}
\begin{tabular}{|c|c|l|l|}
\hline
Prod. & Sample\,\# & $N_{Ce^{3+}}$ & Lab.\\
\hline
6 &	3 & 0.8835 & NIMP\\
6 &	4 & 0.5733 & NIMP\\
6 &	5 & 0.4932 & NIMP\\
7 &	1 & 0.5195 & CERN\\
7 &	2 & 0.5799 & CERN\\
7 & 3 & 0.5386 & NIMP\\
8 &	1 & 0.8030 & CERN\\
8 &	2 & 0.5434 & CERN\\
8 &	3 & 0.4948 & NIMP\\
8 &	4 & 0.5140 & NIMP\\
9 &	1 & 0.9132 & CERN\\
9 &	2 & 1.0730 & CERN\\
9 &	3 & 0.6914 & NIMP\\
9 &	4 & 0.7214 & NIMP\\
10 & 1 & 0.4885 & NIMP\\
11 & 1 & 1.0490 &	NIMP\\
11 & 2 & 0.8990 &	NIMP\\
12 & 1 & 0.8548 & NIMP\\
12 & 2 & 0.9264 & NIMP\\
 &  &  & \\
\hline
\end{tabular}
\end{minipage}
\caption{$Ce^{3+}$ relative concentration ($N_{Ce^{3+}}$) reported per crystal sample. The uncertainty of the $N_{Ce^{3+}}$ corresponds to the stability of the fit procedure (6\,\%). In the last column of the table, information about the laboratory in which the measurement was performed is also given.}
\label{tab:ce_content}
\end{table}


The reproducibility of the transmission spectrum measurement was evaluated repeating the measurement of the same kind of spectrum (along $w$ or $t$) several times and it was found to be within $1\,\%$. The overall measurement process stability, depending on the reliability of the fit function, was evaluated at the level of 6\,\% using the ${N_{Ce^{3+}}}_{w,t}$ values obtained for crystals for which both the transverse spectra were available. 
In particular, it corresponds to the standard deviation of the distribution of ${N_{Ce^{3+}}}_{w,t}$ divided by the corresponding average value over the two transverse spectra $<N_{Ce^{3+}}>$.




$N_{Ce^{3+}}$, as calculated from the absorption spectra, is expected to be correlated with the light yield and the scintillation kinetics expressing the characteristic decay time of the crystals. However, these parameters depend on many other factors that may alter their direct correlation with the concentration of Cerium in the crystal. Possible correlations between the relative concentration of $Ce^{3+}$ absorbing centers, $N_{Ce^{3+}}$, calculated from the absorption spectra and scintillation parameters have been studied and the results are discussed in Sec.~\ref{SS:4-3}.

\label{SS:3-1}
\subsection{Photoluminescence}
\begin{figure}[!htp!]
\begin{center}
\includegraphics[width = 0.70 \textwidth, height=0.50 \textwidth]{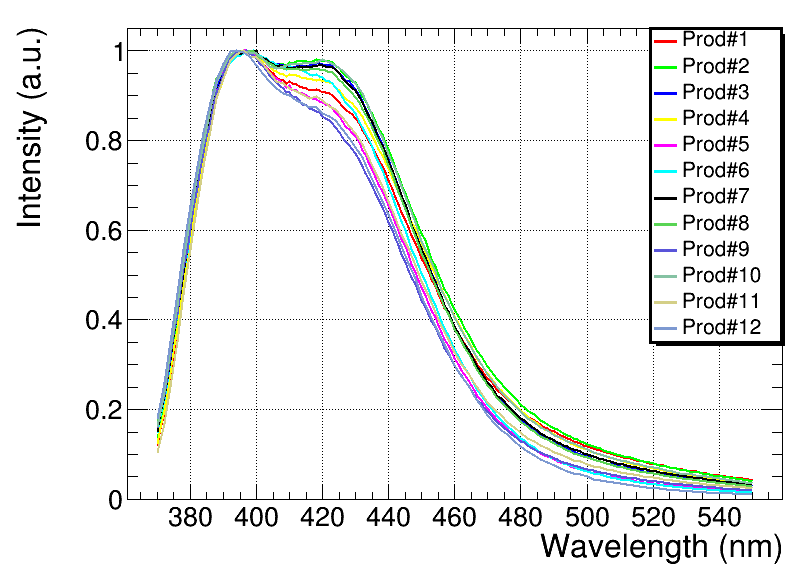}
\caption{Emission spectra for different crystals of different producers ($\lambda_{ex}$ = 358\,nm).}
\label{fig:Luminescence}
\end{center}
\end{figure}

Photoluminescence (PL) measurements were performed for crystals of different producers using an Edimburgh Instruments FS5 Spectrofluorometer at ENEA Casaccia R.C.(Calliope facility lab) in the excitation range 240\,-\,390\,nm and emission range 370\,-\,550\,nm. For the topics of interest in this article, only the emission spectra recorded  in the range 370\,-\,550\,nm by exciting the crystals with $\lambda_{ex}$ = 358\,nm are reported. All the measurements were performed with 2\,nm steps. The emission spectrum measurement reproducibility was found to be 1\,\%. The emission spectra for crystals from different producers, normalized to the maximum intensity value, are given in Fig.~\ref{fig:Luminescence}.


\begin{figure}[!htp!]
\begin{center}
\includegraphics[width = 0.70 \textwidth, height=0.50 \textwidth]{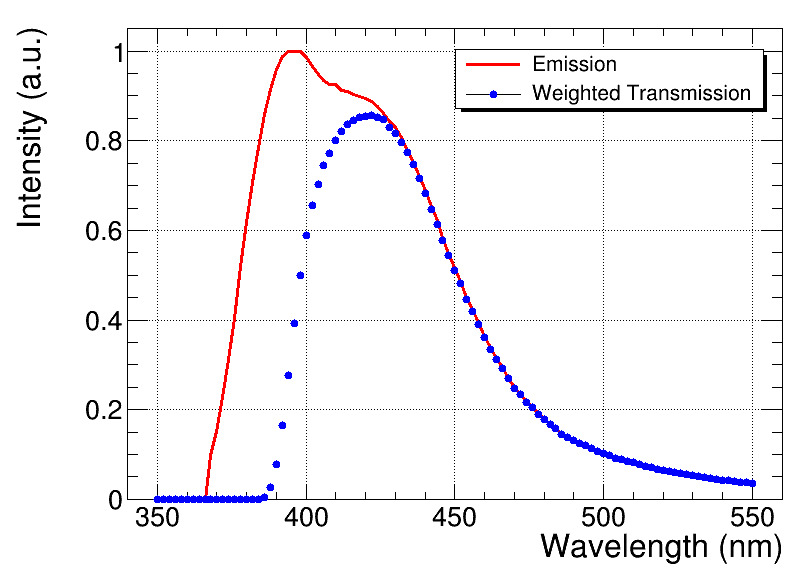}
\caption{Crystal \#66, emission spectrum weighted for  the  transmittance  (blue  dots) and emission spectrum (red line).}
\label{fig:LumiNorm}
\end{center}
\end{figure}

The emission spectra have the same characteristic shape with two peaks at 420\,nm and 396\,nm for all the crystals but the ratio of the two peaks is quite different from one producer to another.

\begin{table}[!htp]
\begin{center}
\begin{tabular}{|c|l|}
\hline
Producer & $I_{420}$/$I_{396}$\\
\hline
1 & 0.98\,$\pm$\,0.01 \\
2 & 0.98\,$\pm$\,0.01 \\
3 & 0.97\,$\pm$\,0.01 \\
4 & 0.93\,$\pm$\,0.01 \\
5 & 0.88\,$\pm$\,0.01 \\
6 & 0.94\,$\pm$\,0.01 \\
7 & 0.97\,$\pm$\,0.01 \\
8 & 0.96\,$\pm$\,0.01 \\
9 & 0.85\,$\pm$\,0.01 \\
10 & 0.98\,$\pm$\,0.01 \\
11 & 0.89\,$\pm$\,0.01 \\
12 & 0.86\,$\pm$\,0.01 \\
\hline
\end{tabular} \\
\caption {Relative emission intensity defined as $I_{420}/I_{396}$ for the crystals studied in the present work.}
\label{tab:RelEmission}
\end{center}
\end{table}In Tab.~\ref{tab:RelEmission}, the relative intensity defined as the ratio of the intensity of the two peaks $I_{420}/I_{396}$ for each producer is reported.
Crystals that exhibit a higher $I_{420}/I_{396}$ ratio have intrinsically a better light collection efficiency (for the same optical quality of the crystal surface and bulk purity). This is due to the smaller presence of the self-absorption mechanism in correspondence of the 420\,nm peak with respect to the 396\,nm peak region \cite{{Chen-2005},{Mao-2008}}.

As example, the emission spectrum weighted for the transmittance is shown in Fig.~\ref{fig:LumiNorm} for the same crystal \#66. The resulting spectrum provides the information necessary to optimize the coupling of the crystals with the light detection sensor.

\label{SS:3-2}
\section{Scintillation properties}
\label{SS:4}
The light output ($LO$) and the decay time ($\tau$) of the crystal samples from each producer were measured with dedicated setup and methods at the the INFN - Sezione di Roma and Sapienza University laboratory (Roma, Italy). The results are shown as the average values over the 15 samples of each producer. Details about the reproducibility of the measurements are provided. 

$LO$ and $\tau$ are key parameters for LYSO:Ce crystal timing applications. The highest possible $LO$ in the shortest possible time frame leads to the best timing performance for which a figure of merit can be defined as the ratio $LO/\tau$. Results for the figure of merit are also shown for all the producers.

Finally, the dependency of $LO$ and $\tau$ on the relative $Ce^{3+}$ concentration has been investigated in Sec.~\ref{SS:4-3} with the aim to explore the possibility to use $Ce^{3+}$ concentration as a quality indicator of the scintillation and timing performance of the crystals.
\begin{figure}[!htp!]
\begin{center}
\includegraphics[width = 0.45 \textwidth, height=0.50 \textwidth]{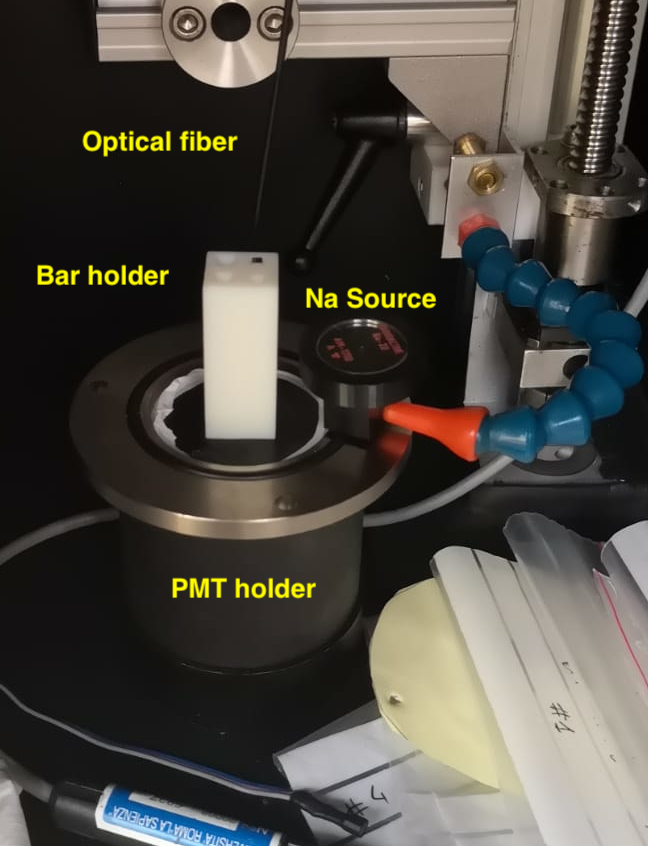}
\caption{Experimental setup used to measure the single crystal bar $LO$ and $\tau$, showing a crystal placed in the plastic holder over the PMT window. The optical fiber coupled to a blue LED and the support for the \ce{^22Na} source are visible too.}
\label{fig:LO_DT_setup}
\end{center}
\end{figure}
\subsection{Experimental setup, methods and tools}
\label{SS:4-1}
\subsubsection*{Setup description}
The experimental setup used for the measurement of the scintillation properties is shown in Fig.~\ref{fig:LO_DT_setup}. It consists of a 51\,mm diameter end window PMT (ET Enterprised model 9256B) placed inside a cylindrical box with a rectangular frame. The frame works as a guide to insert the bar holder which keeps the crystal bar vertical on the PMT photocathode window and is equipped with different transverse section holes for the housing of the 3 bar types. The crystal bars are inserted into the holder without any wrapping. One crystal end face is in contact with the PMT window while the other one is free and in contact with air.
No grease is applied to enhance the PMT-crystal optical contact. This precaution was taken to optimize the reproducibility of the measurement. The setup is enclosed in a black painted box whose temperature is kept stable at 20$^\circ$C (within 0.1-0.2 $^\circ$C over 24\,h) by the use of a chiller. The PMT signal is readout by the DRS4 evaluation board \cite{DRS4}, working at a sampling rate of 2\,GS/s; this allows an integration window for the PMT signal extending up to 500\,ns. 
The single photoelectron (SPE) response is calibrated using a pulsed, fast, blue LED. The LED light is brought inside the box using an optical fiber. 
\subsubsection*{Light output measurement}
The absolute $LO$ measurement is performed using one of the annihilation photons emitted by a \ce{^22Na} radioactive source placed beside the bar and evaluating the position of the 511\,keV photoelectron peak in the crystal signal. The charge of the photoelectron peak is then divided by the SPE charge and by the energy of the photon to obtain the $LO$ value expressed in photoelectrons per MeV of deposited energy. 

\begin{figure}[!ht]
\begin{center}
\includegraphics[width = 0.70 \textwidth, height=0.50 \textwidth]{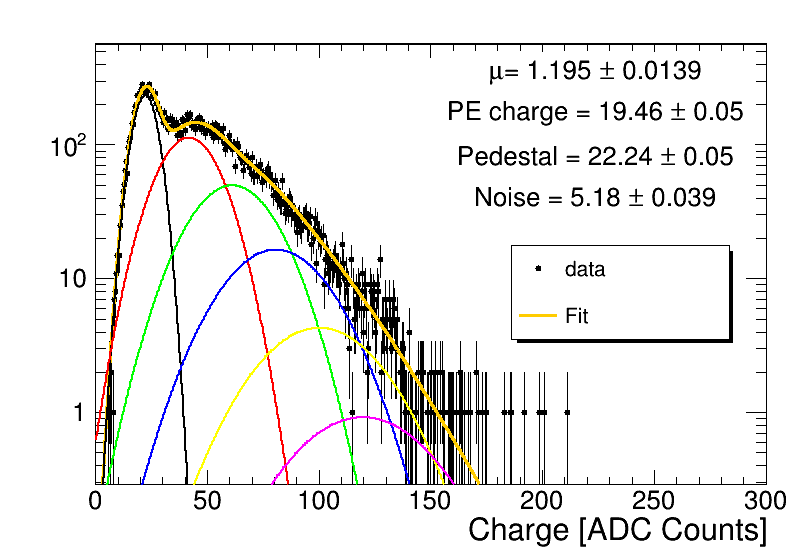}
\includegraphics[width = 0.70 \textwidth, height=0.50 \textwidth]{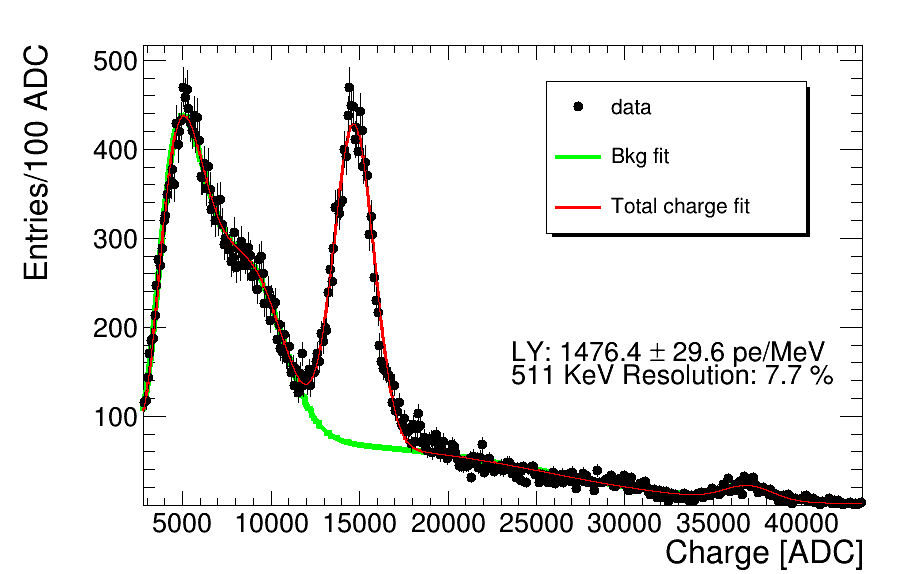}
\caption{Example of charge spectrum used to extract the charge associated to the SPE. The contributes to the fitting function (orange line) from the pedestal (black dotted line), 1 ( red line), 2 (green line), 3 (blue line), 4 (yellow line) and 5 (magenta line) photoelectrons are visible (top). Example of a charge spectrum obtained with the \ce{^22Na} radioactive source.  The total charge fitting function (red line) used to extract the 511\,keV photo-peak and the background events fitting function (green line) are shown (bottom).}
\label{fig:charge_spectra_fit}
\end{center}
\end{figure}

\vspace{2mm}
\textit{SPE}. The SPE charge value is extracted by fitting the charge spectra obtained with the LED with the convolution of a Poisson (accounting for gamma conversion process and first dynode photoelectron collection) and a Gauss distribution (accounting for multiplicative dynode system response), as shown in Fig.~\ref{fig:charge_spectra_fit}, top. In order to improve the fit stability, 5 charge spectra obtained with increasing LED pulses of different intensities are collected and simultaneously fitted leaving the SPE charge as common free parameter.

The PMT signal acquisition is triggered by the coincidence signal provided by the LED driver and the charge is integrated in a 30\,ns window after the baseline subtraction.

\vspace{2mm}
\textit{511\,keV photo-peak}. The charge associated to the 511\,keV photo-peak is obtained using a 17 parameter fit which fully describes the energy deposit of both the 511\,keV and the 1275\,keV photons emitted by \ce{^22Na}, including the contributions due to Compton, photo-electric and back-scatter interactions. A turn-on function is also used to describe the trigger behavior.

In this case the PMT signal acquisition is triggered on the PMT signal itself using an optimal threshold. The charge is integrated in a 450\,ns time window after the baseline subtraction.
 An example of charge spectra used to extract the 511\,keV photo-peak values is presented together with the corresponding fitting functions in Fig.~\ref{fig:charge_spectra_fit}, bottom. 
\subsubsection*{Decay time measurement}
\begin{figure}[!htp!]
\begin{center}
\includegraphics[width = 0.70 \textwidth,height=0.50 \textwidth]{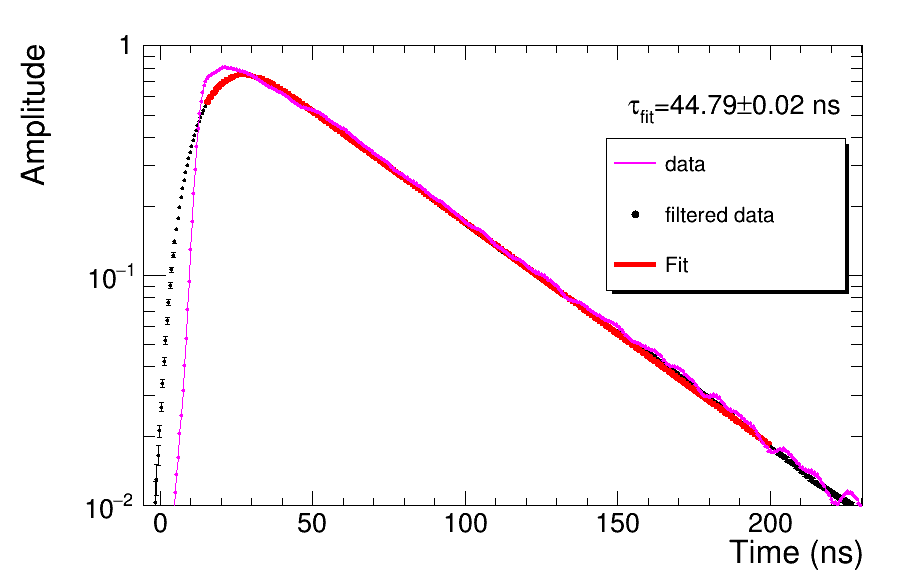}
\caption{Average waveform from a \ce{^22Na} run with the superimposed fit (red line) performed on filtered data (black dots) to estimate the $\tau$ of the crystal.}
\label{fig:fitWaveform}
\end{center}
\end{figure}

The acquisition with a fast sampling digitizer allows the extraction of the scintillation $\tau$ directly from the acquired waveform of the PMT signal. An average over all PMT signals with an associated total charge above roughly 100\,keV in the \ce{^22Na} runs is performed. The average waveform is passed through a Butterworth filter with a cut-off frequency of 20\,MHz to reduce oscillations due the imperfect impedance matching between the PMT anode output and the DRS4 buffer input. $\tau$ is extracted from a fit which includes a single exponential decay function and a Gaussian turn-on. An example of this fit is shown in Fig.~\ref{fig:fitWaveform}. From the average waveform it is also possible to estimate the amount of light emitted in a time window smaller than 450\,ns, integrating the waveform in different time windows.

The reproducibility of the $LO$ and $\tau$ measurements was estimated repeating them daily over one month using a reference crystal and it was found to be 4\,\% and better than 1\,\%, respectively.
\subsection{Measurement results}
\label{SS:4-2}

\begin{figure}[!htp!]
\begin{center}
\includegraphics[width = 0.70 \textwidth, height=0.50 \textwidth]{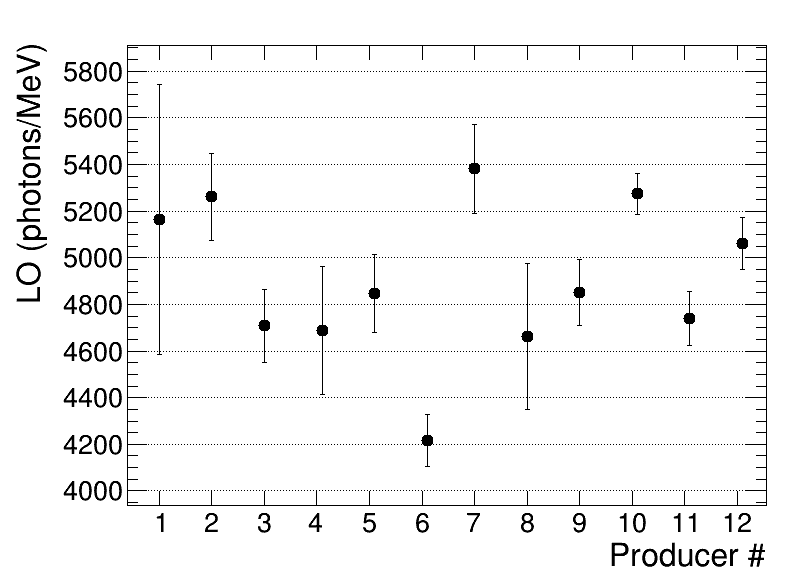}
\includegraphics[width = 0.70 \textwidth, height=0.50 \textwidth]{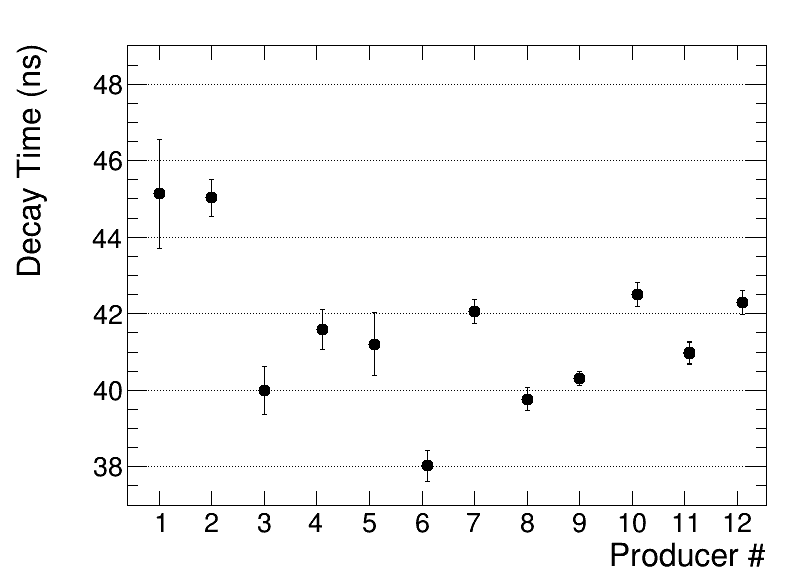}
\caption{$LO$ (top) and $\tau$ (bottom) results for the 12 producers.}
\label{fig:LY_DT_results}
\end{center}
\end{figure}
\begin{figure}[!htp!]
\begin{center}
\includegraphics[width = 0.70 \textwidth,height=0.50 \textwidth]{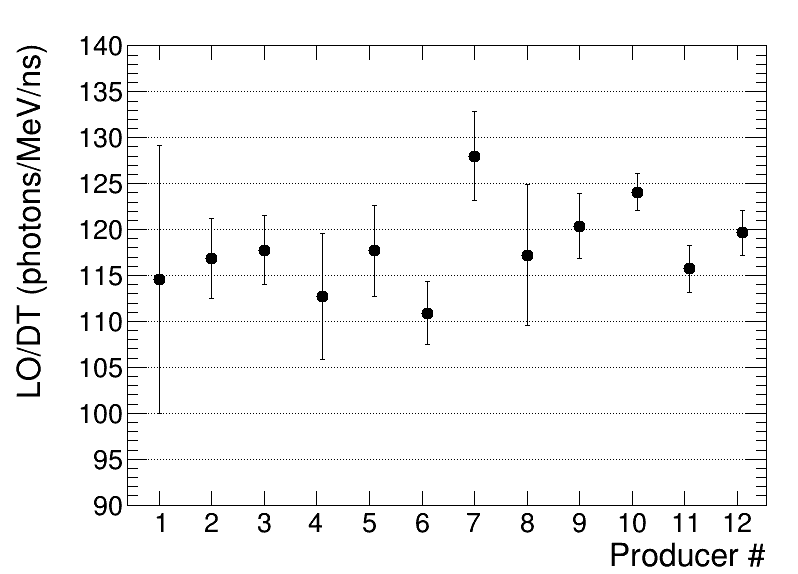}
\caption{Figure of merit for the timing performance of the LYSO:Ce crystals defined as the ratio of $LO$ over $\tau$ .}
\label{fig:fig_of_merit}
\end{center}
\end{figure}
The $LO$ and $\tau$ measurement results are averaged over the 15 crystals provided by each producer and are displayed in Fig.~\ref{fig:LY_DT_results}.
The $LO$ (Fig.~\ref{fig:LY_DT_results}, top) is expressed in photons/MeV and represents the number of scintillation photons produced per MeV of energy deposit which impinge on the photosensor and are successfully detected. It is corrected for the quantum efficiency of the sensor and corresponds to the intrinsic crystal light yield (LY) times the light collection efficiency (LCE). The latter depends on the optical surface quality of the crystal and the transparency of the bulk as well as the crystal-sensor coupling (which is, however, the same for all the crystals).
The quantum efficiency correction factor is obtained by the quantum efficiency of the PMT, as provided by the producer, weighted over the LYSO spectrum and corresponds to about 25\,\%. 

The relative standard deviation of the $LO$ values for different producers is about 8\,\%. The $LO$ standard deviation (error bars in Fig.~\ref{fig:LY_DT_results}, top) for samples of the same producer is mostly comparable with the reproducibility of the measurement (4\,\%), although some show higher values revealing a less uniform $LO$ among the provided samples. The standard deviation value of producer 1 can be explained by 2 outlier crystals.

The $\tau$ value ranges from 45 down to 38\,ns for the slowest to the fastest crystal, as illustrated in Fig.~\ref{fig:LY_DT_results}, bottom. The relative standard deviation of the $\tau$ values for different producers is about 5\,\% while the relative standard deviation for crystals from the same producer is around 1\,\% and thus comparable with the reproducibility of the measurement. Finally, Fig.~\ref{fig:fig_of_merit} shows the figure of merit $LO/\tau$ for the timing performance for each producer. The relative standard deviation of the values from different producers is within $5\,\%$.

\subsection{Study of the main scintillation parameters as a function of $Ce^{3+}$ relative concentration}
\label{SS:4-3}
\begin{figure}[!ht]
\begin{center}
\includegraphics[width = 0.70 \textwidth, height=0.50 \textwidth]{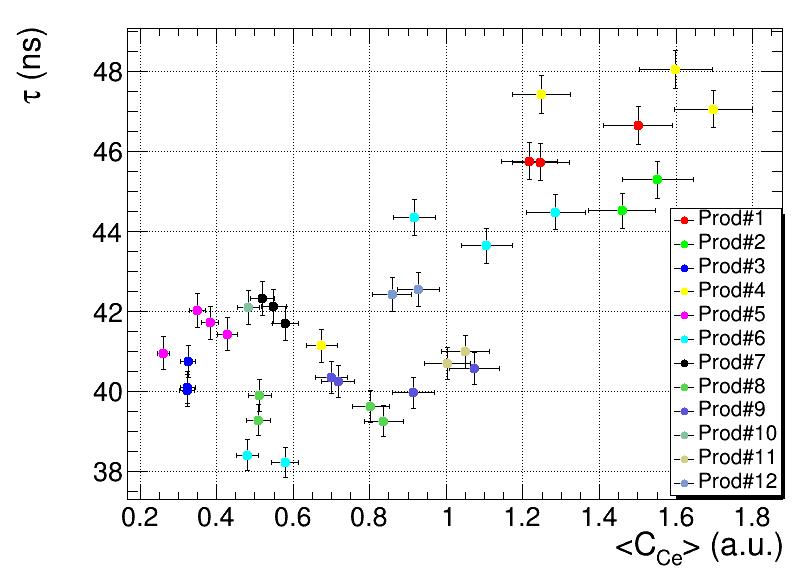}
\caption{$\tau$ as a function of the relative concentration of $Ce^{3+}$ absorbing centers.}
\label{fig:CeVsDt}
\end{center}
\end{figure}

\begin{figure}[!ht]
\begin{center}
\includegraphics[width = 0.70 \textwidth, height=0.50 \textwidth]{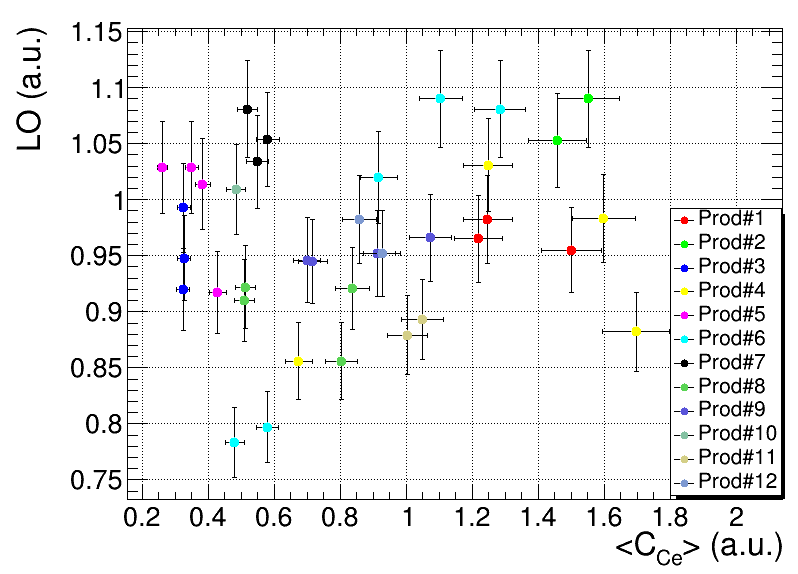}
\caption{$LO$ as a function of the relative concentration of $Ce^{3+}$ absorbing centers. The $LO$ value is normalized to the $LO$ of a reference crystal.}
\label{fig:CeVsLY}
\end{center}
\end{figure}

Figures \ref{fig:CeVsDt} and \ref{fig:CeVsLY} show $\tau$ and $LO$ as a function of the $Ce^{3+}$ relative concentration calculated as described in Sec.~\ref{SS:3-1}. The linear dependence between $\tau$ and $LO$ and the calculated $Ce^{3+}$ relative concentration is too weak to recommend the use of the latter for an indirect assessment of the first two parameters i.e. the scintillation performance of crystals.
In particular, while the $\tau$ trend with respect to $Ce^{3+}$ relative concentration is close to the expectation of linear correlation, for the $LO$ the linear dependency hypothesis is weaker.

The scintillation performance of LYSO:Ce crystals depends on several factors, not only the intensity of the optical absorption peak at 360\,nm, which represents the standard indicator for the concentration of $Ce^{3+}$ absorbing centers (i.e. concentration of Cerium used as scintillation activator). First of all, the intensity of the absorption peak at 360\,nm reflects only the content of $Ce^{3+}$ \cite{ma4071224,MARTINS2017163} while the content of $Ce^{4+}$, which has an important contribution to the scintillation $LO$, remains unknown. The ratio between the $Ce^{3+}$ and $Ce^{4+}$ concentrations depends on possible co-doping applied by each crystal producer and also on unintentional impurities and defects. 

As mentioned above, the samples are too thick for measuring the broad band in the UV region which may possibly give a hint on the $Ce^{3+}/Ce^{4+}$ concentrations ratio. Furthermore, the light yield depends on the competition between radiative and non-radiative recombination. This competition might be strongly affected by co-doping and unintentional impurities due to different raw materials used by different crystal producers.

On the basis of these arguments, the results in Fig.~\ref{fig:CeVsDt} and \ref{fig:CeVsLY} can be explained as a milder sensitivity of $\tau$ to the presence of other dopants and impurities or defects with respect to the one exhibited by the $LO$. Co-doping, defects and impurities depend indeed by the specific LYSO:Ce recipe and growing process chosen by each manufacturer.

\section{$\gamma$ radiation hardness} 
\label{SS:5}
Radiation hardness of the crystal samples against ionizing radiation by $\gamma$ rays was studied at the Calliope facility of ENEA-Casaccia Research Centre (Rome, Italy). Calliope is a pool-type facility equipped with a $^{60}$Co radio-isotopic source array in a large volume shielded cell \cite{Calliope}. The irradiation tests involved at least one crystal bar of type 2 for each producer.  All the samples were irradiated at the same dose rate of 9\,kGy/h and received a total integrated absorbed dose of 50\,kGy. The dose rate value is experimentally measured by an alanine-ESR dosimetric system mapping the Calliope irradiation area. The dose rate uncertainty is 5\,\%.
\subsection{Scintillation properties}
\label{SS:5-1}
All the irradiated samples were measured before and after the irradiation with the setup described in Sec.~\ref{SS:4-1}.
\begin{figure}[!htp!]
\begin{center}
\includegraphics[width = 0.70 \textwidth,height=0.50 \textwidth]{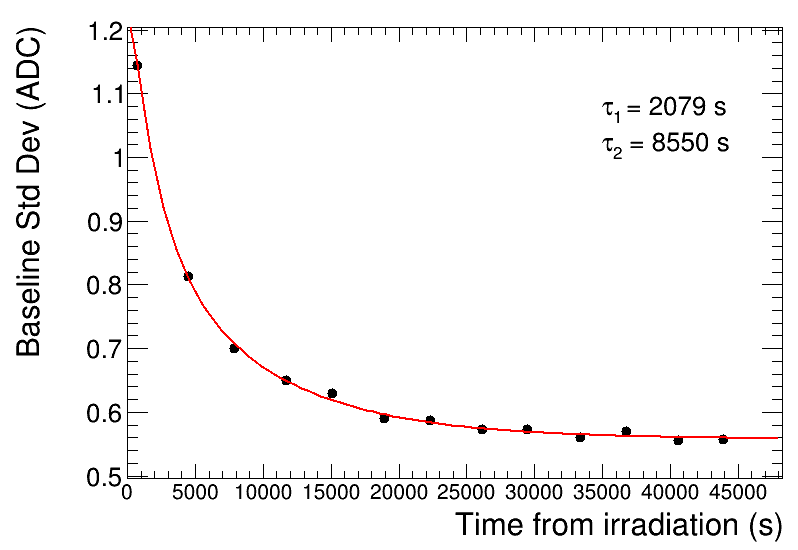}
\caption{Impact of the LYSO phosphorescence light on the standard deviation of the PMT signal baseline as a function of the time from the end of the irradiation. The standard deviation of the PMT signal charge is calculated in a 20\,ns time window before the scintillation signal and averaged over the events of a source run.}
\label{fig:afterglow}
\end{center}
\end{figure}

\begin{figure}[!htpp]
\begin{center}
\includegraphics[width = 0.70 \textwidth, height=0.50 \textwidth]{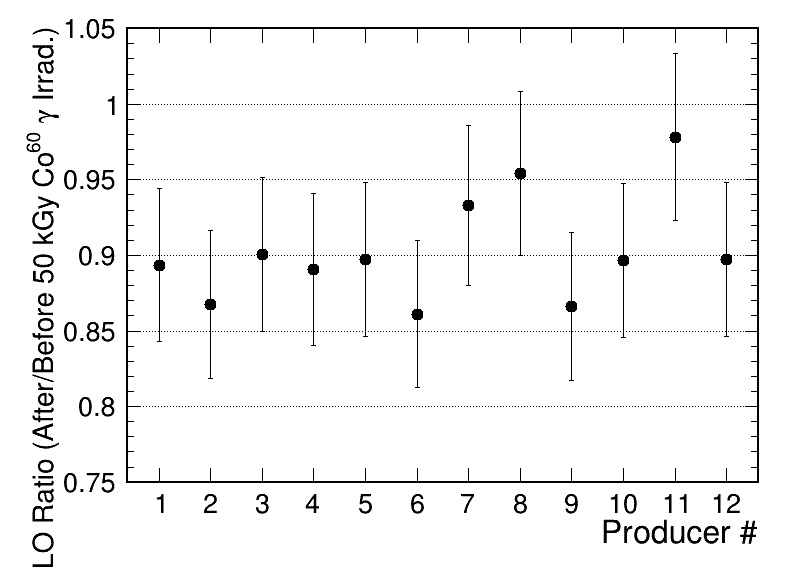}
\includegraphics[width = 0.70 \textwidth, height=0.50 \textwidth]{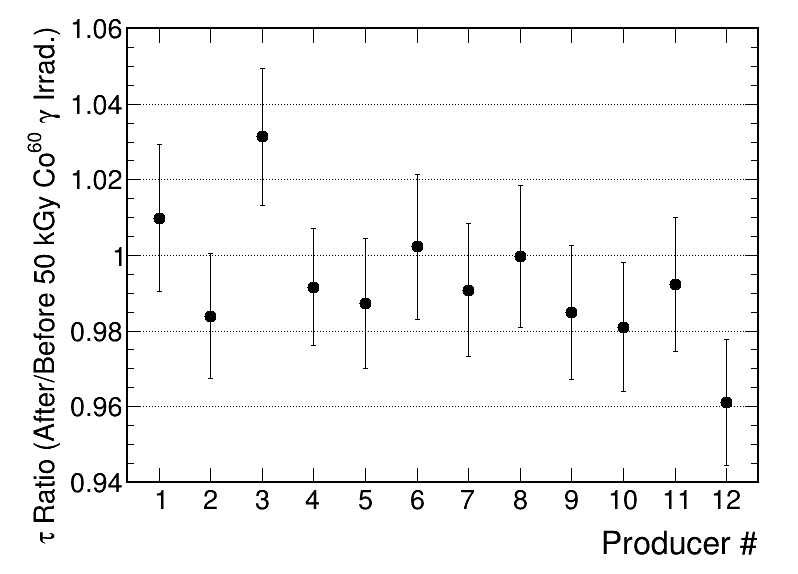}
\caption{Ratio of $LO$ (top) and $\tau$ (bottom) after and before $\gamma$ irradiation for the 12 producers. The error bars are determined by propagation of the measurement uncertainties corresponding in this case to the reproducibility of the $LO$ and $\tau$ measurements.}
\label{fig:LY_DT_after_irrresults}
\end{center}
\end{figure}

After irradiation, all the crystals exhibited phosphorescence light with an approximate decay time of 2\,-\,3\,h as estimated from the presence of a transient noise in the baseline of the PMT signal acquired $\sim$\,every hour for 12\,h and displayed in Fig.~\ref{fig:afterglow}.
For this reason, the samples were measured again at least 16\,h after the irradiation to evaluate the ratio of the $LO$ and the $\tau$ after and before irradiation. The results are shown in Fig.~\ref{fig:LY_DT_after_irrresults}. The average light output loss amounts to $9\,\%$ with a relative standard deviation of $3\,\%$ among the different producers (Fig.~\ref{fig:LY_DT_after_irrresults}, top).
 The scintillation $\tau$ (Fig.~\ref{fig:LY_DT_after_irrresults}, bottom) after irradiation remains unchanged within the measurement uncertainties compared to the pre-irradiation value for most of the producers. The average ratio of $\tau$ after and before the irradiation is $1\,\%$ with a standard deviation of $2\,\%$.

In general, the scintillation mechanism of LYSO:Ce is not damaged by $\gamma$-ray irradiation \cite{Mao2009GammaRI}. The $LO$ decrease depends on the $\gamma$-induced transparency loss which is due to the creation of absorbing centers. The $LO$ can be further recovered through a air annealing of the crystal at $\sim$300$^\circ$C for some hours. Slow (few days) spontaneous recovery can also be observed at room temperature \cite{ma4071224}.
\clearpage
\label{SS:5-2}
\section{Scintillation properties at low temperature}
\label{SS:6}
Due to its radiation hardness against photons and hadrons, LYSO:Ce can be employed for timing purposes in the harsh environment of the new generation particle colliders such as the HL-LHC. Here, to mitigate the impact of the radiation damage on the performance of the detector components, especially the silicon ones, the operating temperature is usually lowered below 0$^\circ$C by some tens of degrees. This will be, for example, the case of the barrel part of the timing detector of CMS-phase II. In BTL, LYSO crystals are coupled to Silicon PhotoMultipliers (SiPM). Radiation exposure increases the noise due to the SiPM dark count rate and lowers the $LO$ of the crystals deteriorating the time resolution. For this reason the detector will be operated at low temperature, between $-45^\circ$C and $-35^\circ$C.

With the aim to extend and complete the set of information collected in this paper, additional measurements of $LO$ and $\tau$ in this range of temperatures for crystal bars from each of the 12 producers were performed. The experimental setup and the results are presented in this section.
\subsection{Experimental setup}
\label{SS:6-1}
The experimental bench used for cold measurements of LYSO:Ce crystal features the same concept of the PMT bench used in the crystal characterization campaign at room temperature (20$^\circ$C) and described in Sec.~\ref{SS:4}. Also the methods and the analysis tools to obtain the values of $LO$ (expressed in photoelectrons per MeV of deposited energy) and $\tau$ are the same. The $LO$ value is corrected for the temperature dependency of the PMT gain using the charge of the SPE measured at the same temperature with the LED. The same LED runs have been used to exclude a non-negligible dependency of the PMT quantum efficiency (QE) on temperature. This was obtained verifying that the average number of photoelectrons in a LED run (LED intensity set to give an average number of photoelectrons $\simeq1$) remains constant with the temperature.
To reach and stabilize the temperature down to $-30^\circ$C, the setup was enclosed into a thermostatic chamber (Angelantoni TY110) and equipped with a temperature monitor. The temperature fluctuations during a standard data taking have been measured and found to be $\pm\,0.2 ^\circ$C. In a preliminary study, the response of the PMT used (Hamamatsu R7378) was measured and proved to be linear down to $-30^\circ$C. 
 The PMT signal is brought outside the chamber through a circular feedthrough and readout by a 12 bit 3.2\,GS/s digitizer (CAEN DT5743).  The reproducibility of the $LO$ and the $\tau$ measurement was evaluated repeating several times the corresponding measurements using a reference crystal and it was found to be 2\,\% and $< $1\,\%, respectively. The better performance in term of $LO$ measurement reproducibility of this test bench with respect to the one used for the measurements described in Sec.~\ref{SS:4-1} is probably due to the better temperature stabilization provided by the high performance thermostatic chamber in which the setup was enclosed. 
\begin{figure}[!ht]
\begin{center}
\includegraphics [width = 1.0 \textwidth] {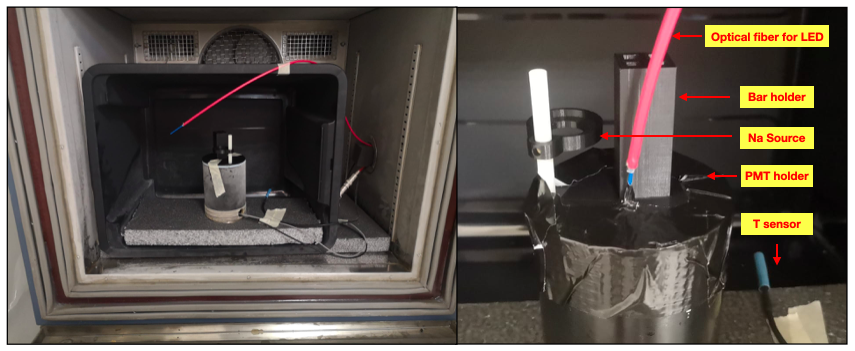}
\caption{(left) A picture of the experimental bench used for the characterization of single crystal bars at cold temperatures. The bench is inserted in a thermostatic chamber able to provide stable temperatures down to $-40^\circ$C. (right) A detailed picture of the setup components.}
\label{fig:labe_setup}
\end{center}
\end{figure}
\subsection{Results}
\label{SS:6-2}
\begin{figure}[!hbt]
\begin{center}
  \includegraphics[width = 0.70 \textwidth, height=0.50 \textwidth]{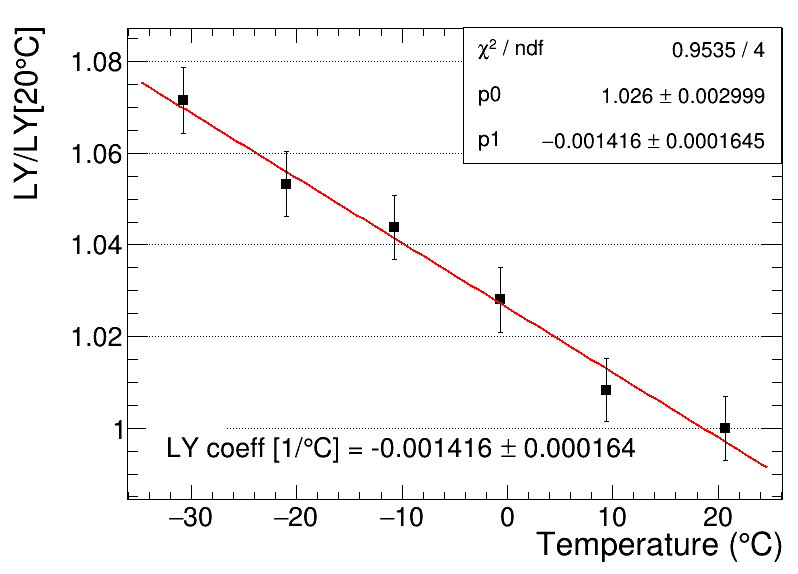}
  \includegraphics[width = 0.70 \textwidth, height=0.50 \textwidth]{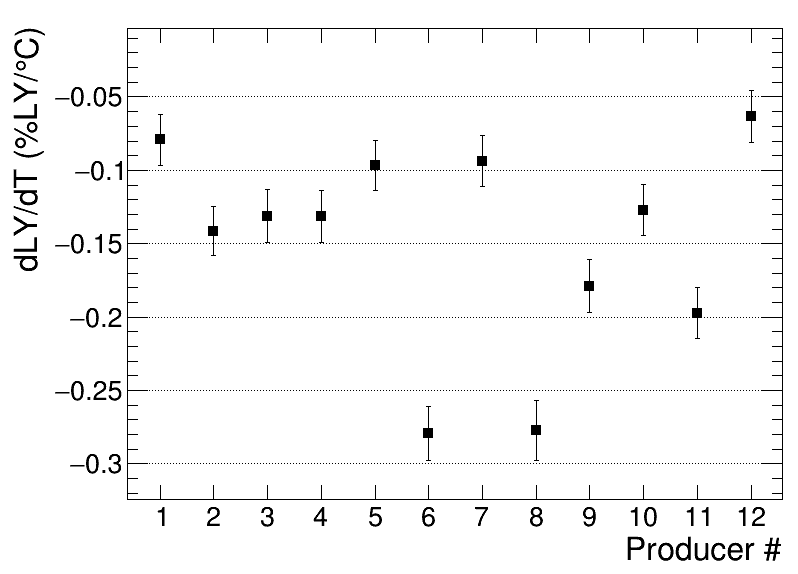}
\caption{(top) LY normalized to the corresponding value at T=\,20$^\circ$C as a function of the temperature. The normalized LY error bars were determined by propagation of the measurement uncertainties.
From the linear fit, the LY temperature coefficient is obtained. (bottom) Light yield temperature coefficient for the 12 producers. The error bars correspond to the fit uncertainties.}
\label{fig:LO}
\end{center}
\end{figure}

At least one crystal bar of the smallest geometry for each of the 12 producers was measured. Six measurement points have been acquired with temperatures ranging from 20$^\circ$C down to $-30^\circ$C. Lowering the temperature, both the $LO$ and $\tau$ increase slowly. In Fig.~\ref{fig:LO} (top) an example of $LO$ as a function of the temperature and normalized to the corresponding value at T=\,20$^\circ$C is shown.

\begin{figure}[!hbt]
\begin{center}
\includegraphics[width = 0.70 \textwidth, height=0.50 \textwidth]{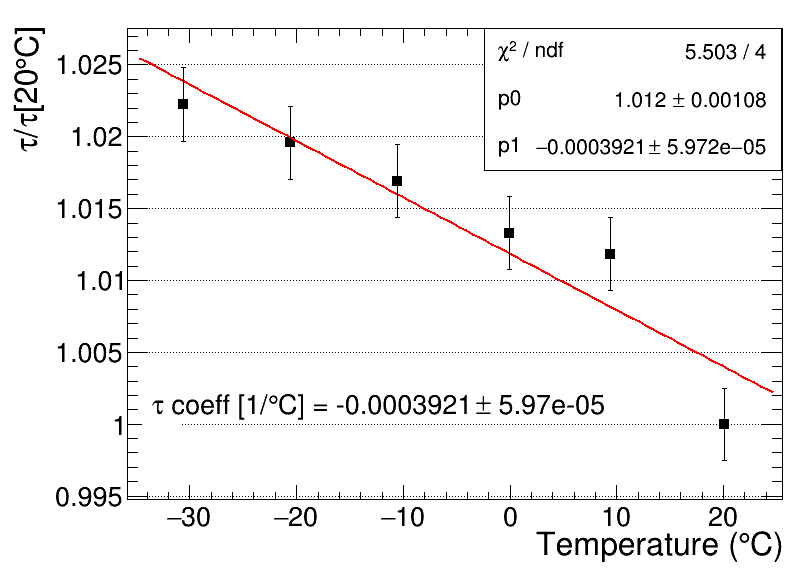}
  \includegraphics[width = 0.70 \textwidth, height=0.50 \textwidth]{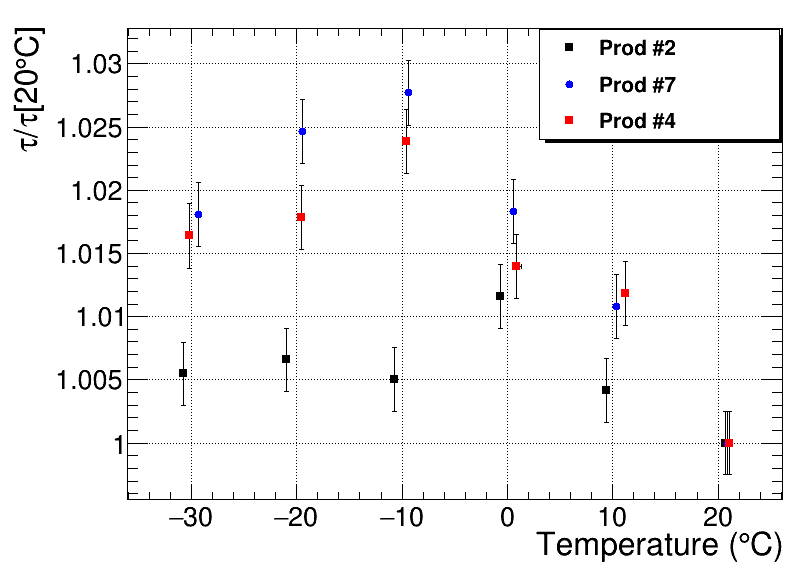}
\caption{(top) $\tau$ normalized to the corresponding value at T=\,20$^\circ$C as a function of the temperature. The normalized $\tau$ error bars correspond to the reproducibility of the $\tau$ measurement.
For this producer, $\tau$ has a linear behavior and from a linear fit the $\tau$ temperature coefficient can be obtained. (bottom) $\tau$ normalized to the corresponding value at T=\,20$^\circ$C as a function of the temperature for producer 2, 4 and 7.
For this producers, $\tau$ does not feature a linear dependency on the temperature.}
\label{fig:DT}
\end{center}
\end{figure}
The $LO$ is linear with the temperature for all producers. The temperature coefficient is on average $-0.15\,\%/^\circ$C ranging between $-0.28\,\%/^\circ$C and $-0.08\,\%/^\circ$C 
as shown in Fig.~\ref{fig:LO} (bottom).
The $LO$ relative variation as a function of the temperature is equal to the light yield (LY) relative variation because the $LO$ can be factorized as $LY \times LCE \times QE$ and the LCE and the QE can be assumed constant with the temperature and therefore cancel out in the ratio.

$\tau$ dependency on the temperature is linear down to $-30^\circ$C only for 6 producers over 12 (regression coefficient R$>$0.85) and in general the variation with temperature is smaller than for the $LO$. In Fig.~\ref{fig:DT} (top) the linear dependency of $\tau$ for producer 5 is shown as an example.
For the other producers, no linear relation between the temperature and $\tau$ can be assumed (R$<$0.75).
In Fig.~\ref{fig:DT} (bottom), $\tau$ vs. T is shown for crystals from this subset of producers; in particular for producer 2 (R=0.41), 4 (R=0.73) and 7 (R=0.76).
For these producers, additional measurement points at low temperature would be needed for a more rigorous description of $\tau$ dependency down to $-30^\circ$C.

\begin{figure}[!hbt]
\begin{center}
  \includegraphics[width = 0.70 \textwidth, height=0.50 \textwidth]{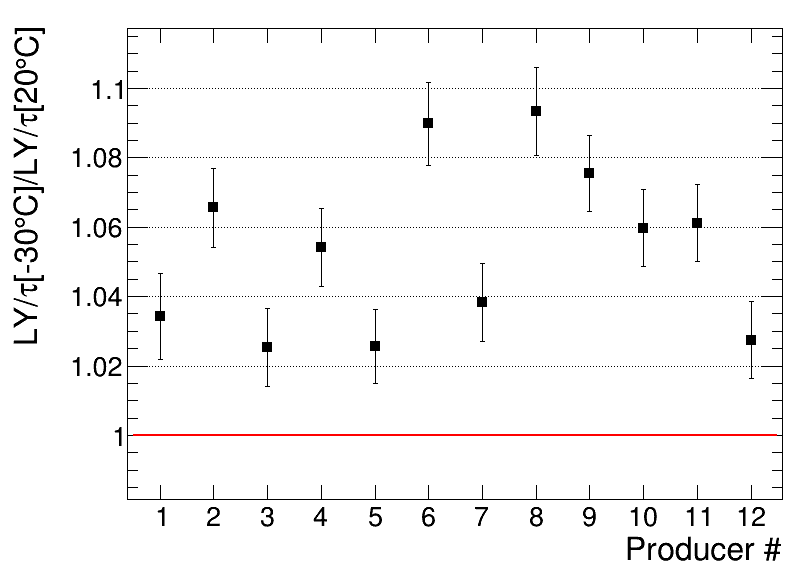}
\caption{Ratio of the figure of merit (here expressed as $LY/\tau$) for timing performance measured at $-30^\circ$C and at 20$^\circ$C for the 12 producers. The error bars were determined by propagation of the measurement uncertainties.}
\label{fig:cold_measurements}
\end{center}
\end{figure}

In Fig.~\ref{fig:cold_measurements} the ratio of the figure of merit ($LO/\tau$) measured at $-30^\circ$C and at 20$^\circ$C is also shown. Its average value and standard deviation are 1.05 and 0.02 respectively. For all producers the ratio is $>$1. This demonstrates that lowering the operating temperature of the crystal can help to improve their timing performance.
\clearpage
\section{Discussion}

A set of 15 small crystal bars ($3\,mm\times3\,mm\times57\,mm $) from 12 different producers were studied and compared with respect to a set of properties and performance fundamental for HEP applications with a special focus on timing applications.

All producers are shown to have mastered the cutting technology producing samples with uniform dimensions at the level of per mille, and within the requested specifications at a level better than 1\,\%.
From the dimensions and the mass measurement, the crystal density value was derived for every sample. It ranges from 7.1 to 7.4\,g/cm$^3$ and its relative standard deviation among the samples of the same producer is well below 1\,\%.

The mass density study is complemented, for at least one crystal per producer, by inductively coupled plasma mass spectrometry (ICP-MS) measurements from which the Yttrium molar fraction was evaluated. The Yttrium fraction is indeed expected to linearly correlate with the mass density. The expectation has been confirmed by data ($R=0.95$) and the spread of the Yttrium fraction among the different producers is about 30\,\%.

Optical transmission spectra and photoluminescence properties were also studied for all producers. In particular, the evaluation of the relative concentration of the main crystal luminescence center ($Ce^{3+}$) was obtained from the transmission spectra. Its correlation with the light output ($LO$) and decay time ($\tau$) of the crystals has been investigated in the attempt to establish a method to characterize the timing performance of the crystals. 
The data do not match the expectations showing a poor linear correlation of the ($Ce^{3+}$) relative concentration with both scintillation parameters. This has been mainly ascribed to the possible presence of different co-dopants, impurities and defects which may have an important role in the scintillation dynamics.

$LO$ and $\tau$ were measured for all the crystal samples, together with the figure of merit for timing application defined as $LO/\tau$. all producers' samples show similar scintillation properties.
The spread of the $LO$ value for different producers is at the level of 8\,\% while for $\tau$, ranging from 38 to 45\,ns, it is within 5\,\%. The uniformity of the crystal samples provided by each producer with respect to these scintillation parameters is comparable with the reproducibility of the measurements: 4\,\% for the $LO$ and 1\,\% for $\tau$.

In order to test the radiation hardness of the crystal samples against $\gamma$, $LO$ and $\tau$ were also measured after irradiation with 50\,kGy at a dose rate of 9\,kGy/h  for a subsample of crystals from all 12 producers. While $\tau$ remains essentially unchanged for all producers, the $LO$ loss is on average at the level of 10\,\%. The study did not include a thermal annealing campaign. Nevertheless it is a well established concept that the $LO$ damage is not permanent and it can be fully recovered by thermal annealing.

Finally, the $LO$ and $\tau$ dependency on temperature was analyzed for a subsample of crystals down to $-30^\circ$C. The $LO$ exhibits a linear dependency on temperature with a temperature coefficient ranging between $-0.28\,\%/^\circ$C and $-0.08\,\%/^\circ$C. Only 6 producer over 12 shows a linear $\tau$ dependency on the temperature down to $-30^\circ$C. More data points at low temperature would be needed to study the non-linearity of $\tau$ for the other producers.

Nevertheless, the figure of merit at $-30^\circ$C compared with the results obtained at $20^\circ$C shows that lowering the operating temperature of the crystals can help to improve their timing performance. This holds true for all the producers and with an relative standard deviation of $\simeq$2\,\%.

The most important crystal features measured in this study are summarized in Tab.~\ref{tab:summary_1} and Tab.~\ref{tab:summary_2} for each producer. All producers showed similar characteristics within $\simeq$10\,\%, except for the $Ce^{3+}$ relative concentration and the LY temperature coefficient. For these crystal properties the spread among the producers is at the level of 50\,\%. Despite this, their impact on the key performance for HEP and especially for timing application is limited. The $Ce^{3+}$ relative concentration has shown a poor correlation with LO and $\tau$ while the spread in the LY temperature coefficients does not reflect in the figure of merit $LY/\tau$.



\begin{table}[H] 
\begin{center}
\begin{tabular}{|c|c|c|c|c|c|c|c|}
\hline
Crystal & M. density & $N_{Ce^{3+}}$ & $LO$ & $\tau$& $LO/\tau$ \\
Prod. & (g/cm$^{3}$) & (a.u) & (ph./MeV) & (ns)& (ph/MeV\,ns) \\
\hline
1 & 7.088\,$\pm$\,0.020 & 1.430\,$\pm$\,0.216 & 5164\,$\pm$\,580 & 45.13\,$\pm$\,1.43 & 115\,$\pm$\,15 \\
2 & 7.093\,$\pm$\,0.008 & 1.704\,$\pm$\,0.283 & 5261\,$\pm$\,186 & 45.03\,$\pm$\,0.49 & 117\,$\pm$\,4 \\
3 & 7.250\,$\pm$\,0.005 & 0.324\,$\pm$\,0.001 & 4708\,$\pm$\,156 & 39.99\,$\pm$\,0.63 & 118\,$\pm$\,4 \\
4 & 7.137\,$\pm$\,0.006 & 1.609\,$\pm$\,0.299 & 4688\,$\pm$\,273 & 41.60\,$\pm$\,0.52 & 113\,$\pm$\,7 \\
5 & 7.103\,$\pm$\,0.008 & 0.327\,$\pm$\,0.052 & 4847\,$\pm$\,169 & 41.21\,$\pm$\,0.93 & 118\,$\pm$\,5 \\
6 & 7.109\,$\pm$\,0.011 & 1.091\,$\pm$\,0.164 & 4216\,$\pm$\,116 & 38.02\,$\pm$\,0.41 & 111\,$\pm$\,3 \\
7 & 7.313\,$\pm$\,0.009 & 0.546\,$\pm$\,0.025 & 5381\,$\pm$\,190 & 42.05\,$\pm$\,0.31 & 128\,$\pm$\,5 \\
8 & 7.175\,$\pm$\,0.008 & 0.589\,$\pm$\,0.125 & 4662\,$\pm$\,313 & 39.76\,$\pm$\,0.30 & 117\,$\pm$\,8 \\
9 & 7.078\,$\pm$\,0.016 & 0.850\,$\pm$\,0.155 & 4852\,$\pm$\,141 & 40.30\,$\pm$\,0.18 & 120\,$\pm$\,3 \\
10 & 7.334\,$\pm$\,0.009 & 0.488\,$\pm$\,0.029 & 5274\,$\pm$\,89 & 42.50\,$\pm$\,0.31 & 124\,$\pm$\,2 \\
11 & 7.116\,$\pm$\,0.006 & 0.974\,$\pm$\,0.075 & 4740\,$\pm$\,116 & 40.96\,$\pm$\,0.30 & 116\,$\pm$\,2 \\
12 & 7.110\,$\pm$\,0.008 & 0.891\,$\pm$\,0.036 & 5061\,$\pm$\,111 & 42.29\,$\pm$\,0.31 & 120\,$\pm$\,2 \\
\hline
\end{tabular}
\caption{Average and standard deviation values for Mass density, $N_{Ce^{3+}}$, $LO$, $\tau$ and $LO/\tau$ measured for the crystals from each producers.}
\label{tab:summary_1}
\end{center}
\end{table}

\begin{table}[H]
\begin{center}
\begin{tabular}{|c|c|c|c|c|c|c|c|}
\hline
Crystal &  $LO_{irr}/LO$ & $\tau_{irr}/\tau$ & $dLY/dT$ & $\tau_{-30^\circ C}$& \textrm{\underline{$(LO/\tau)_{-30^\circ C}$}} \\
Prod. &  &  & ($\%LY/^\circ$C) & (ns)& $(LO/\tau)_{20^\circ C}$  \\
\hline
1 & 0.893\,$\pm$\,0.050 & 1.009\,$\pm$\,0.019 & -0.079\,$\pm$\,0.017 & 46.28\,$\pm$\,1.12 & 1.034\,$\pm$\,0.012 \\
2 & 0.868\,$\pm$\,0.049 & 0.984\,$\pm$\,0.016 & -0.141\,$\pm$\,0.017 & 45.83\,$\pm$\,0.11 & 1.065\,$\pm$\,0.011 \\
3 & 0.900\,$\pm$\,0.051 & 1.031\,$\pm$\,0.018 & -0.131\,$\pm$\,0.018 & 43.86\,$\pm$\,0.11 & 1.025\,$\pm$\,0.011 \\
4 & 0.891\,$\pm$\,0.050 & 0.992\,$\pm$\,0.015 & -0.131\,$\pm$\,0.018 & 42.18\,$\pm$\,0.10 & 1.054\,$\pm$\,0.011 \\
5 & 0.897\,$\pm$\,0.051 & 0.987\,$\pm$\,0.017 & -0.097\,$\pm$\,0.017 & 42.29\,$\pm$\,0.11 & 1.026\,$\pm$\,0.011 \\
6 & 0.861\,$\pm$\,0.049 & 1.002\,$\pm$\,0.019 & -0.279\,$\pm$\,0.018 & 38.67\,$\pm$\,0.10 & 1.090\,$\pm$\,0.012 \\
7 & 0.933\,$\pm$\,0.053 & 0.991\,$\pm$\,0.018 & -0.094\,$\pm$\,0.017 & 43.35\,$\pm$\,0.11 & 1.038\,$\pm$\,0.011 \\
8 & 0.954\,$\pm$\,0.054 & 1.000\,$\pm$\,0.019 & -0.277\,$\pm$\,0.020 & 42.44\,$\pm$\,0.11 & 1.093\,$\pm$\,0.013 \\
9 & 0.866\,$\pm$\,0.049 & 0.985\,$\pm$\,0.018 & -0.179\,$\pm$\,0.018 & 41.59\,$\pm$\,0.10 & 1.075\,$\pm$\,0.011 \\
10 & 0.897\,$\pm$\,0.051 & 0.981\,$\pm$\,0.017 & -0.127\,$\pm$\,0.018 & 43.49\,$\pm$\,0.11 & 1.060\,$\pm$\,0.011 \\
11 & 0.978\,$\pm$\,0.055 & 0.992\,$\pm$\,0.018 & -0.197\,$\pm$\,0.017 & 42.23\,$\pm$\,0.11 & 1.061\,$\pm$\,0.011 \\
12 & 0.897\,$\pm$\,0.051 & 0.961\,$\pm$\,0.017 & -0.063\,$\pm$\,0.018 & 43.44\,$\pm$\,0.11 & 1.027\,$\pm$\,0.011 \\
\hline
\end{tabular}
\caption{Summary of the crystal scintillation properties measured after $\gamma$ irradiation and at low temperatures down to $-30^\circ$C for at least a crystal per producer.}
\label{tab:summary_2}
\end{center}
\end{table}

\clearpage
\section{Conclusions}
A comprehensive and systematic study of LYSO:Ce ($[Lu_{(1-x)}Y_x]_2SiO_5$:$Ce$) crystals is presented. It involves for the first time a large number of crystal samples (180) of the same size from several producers. 
The study consists of a comparative characterization of LYSO:Ce crystal products available on the market and aims, in particular, to investigate key parameters of timing applications for HEP. 

A set of 15 small crystal bars ($3\,mm\times3\,mm\times57\,mm $) for each producer were measured with respect to mechanical, optical and scintillation properties. The latter were studied before and after the irradiation of the crystals with a 50\,kGy integrated dose of $\gamma$-ray and at temperatures down to $-30^\circ$C.  
The timing performance of the crystals was evaluated by a figure of merit defined as $LO/\tau$. 
Finally, the number of the samples provided by each producer allowed for the study of the uniformity of the crystal properties within a producer batch.

The LYSO:Ce products considered in this study fully qualify for timing applications at future HEP colliders. LYSO:Ce crystals of all producers show in general similar properties and an excellent uniformity of the samples. The spread of the crystal characteristics with a direct impact on the timing performance is within 10\,\% among the different producers.

This review of LYSO:Ce crystals does not identify a single producer or a set of producers with globally superior performance. The detected differences in the crystal products, although limited, could however be used to guide the selection process of the LYSO:Ce crystals best suited for a specific application.


\begin{appendix}
\section{Appendix A: Absorbance analytical expression in the approximation of multiple reflection between parallel crystal faces}
\label{A}

The absorbance is defined as:
\begin{equation} \label{eq1}
A=2-log_{10}\,T(\%)
\end{equation}
where $T$ corresponds, in the present study, to the measured optical transmission (transmittance). 
\begin{figure}[!ht]
\begin{center}
\includegraphics [width = 1.0 \textwidth] {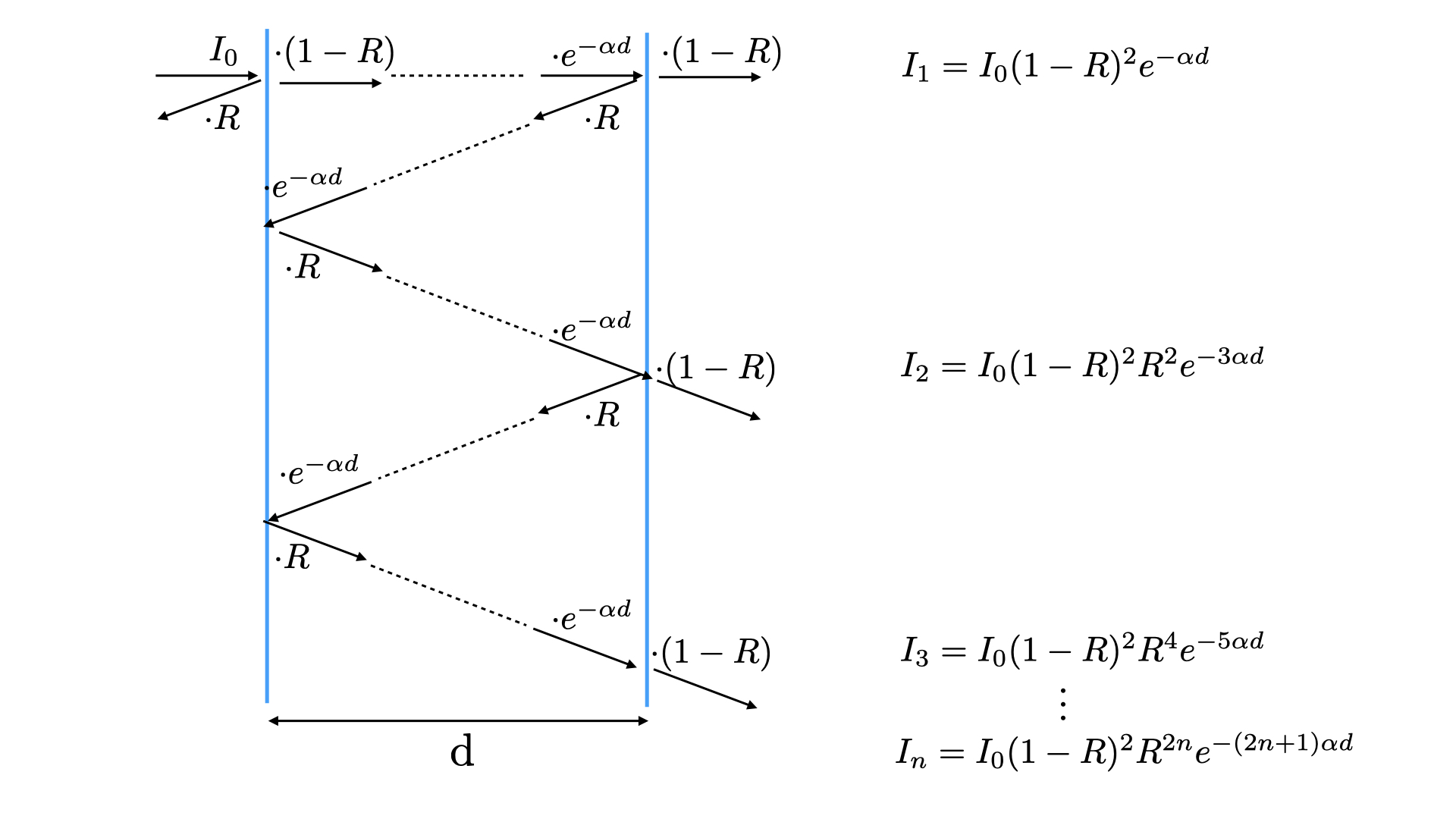}
\caption{Sketch of the multiple reflection of the light between parallel crystal faces at a distance d. The analytical expressions for the contributions of the reflected, absorbed and transmitted light are also reported along the light path.}
\label{scheme}
\end{center}
\end{figure}

The transmittance is defined as the ratio $I/I_{0}$ of light intensities at the exit ($I$) and the entrance ($I_{0}$) of the measured sample. When accounting for multiple reflections on the crystal faces, the numerator is given by the sum of the $I_j$ contributions exiting the crystal and displayed in Fig.~\ref{scheme}:


\begin{equation} \label{eq1A}
T=\frac{I}{{I_0}}=\lim_{n \to \infty}{\frac{\sum_{j=1}^n I_j}{I_0}}= \lim_{n \to \infty}{(1-R)^2 e^{-\alpha d}[1+R^2 e^{-2\alpha d}+R^4 e^{-4\alpha d}+...R^{2n} e^{-2n\alpha d}]}
\end{equation}
\noindent where $R$ and $\alpha$ are the reflection and the absorption coefficients, respectively, and $d$ is the sample transverse size ($w$ or $t$).

Since the term in square bracket corresponds to a geometric progression with common ratio $R^2 e^{-2\alpha d}$, T can be written as:
\begin{equation} \label{eq2A}
T=(1-R)^2 e^{-\alpha d}\lim_{n \to \infty}{\frac{1-(R^2 e^{-2\alpha d})^n}{1-R^2 e^{-2\alpha d}}}
\end{equation}
\noindent Considering the value of $\alpha$ and the value of the refraction index of LYSO ($n_r\simeq 1.7$) in the analyzed ROI as reported in \cite{appendix_paper} and the expression of $R$ at normal incidence:
\begin{equation} \label{eq3A}
R=(\frac{n-1}{n+1})^2
\end{equation}
\noindent $R^2 e^{-2\alpha d}<<1$ holds true and T converge to:
\begin{equation} \label{eq4A}
T=(1-R)^2 e^{-\alpha d}\frac{1}{1-R^2 e^{-2\alpha d}}\simeq (1-R)^2 e^{-\alpha d}
\end{equation}
\noindent Using Eq.~\ref{eq4A} in Eq.~\ref{eq1} and $R$ being constant in the considered ROI, A is found to be proportional to d: 
\begin{equation} \label{eq7}
A \sim\alpha \cdot d
\end{equation}

\end{appendix}

\section*{Acknowledgement}
The authors would like to thank Etiennette Auffray who made available the spectrometer at CERN used for transmission measurements,
Massimo Nuccetelli, Antonio Zullo, Maurizio Zullo, Angelo Mattei and Marco Iannone (INFN\,-\,Rome) for supporting mechanical design and production of the experimental test benches, Luigi Recchia and Antonio Girardi (INFN\,-\,Rome) for
supporting the readout electronics of the test benches.
Lastly, the authors thank Sapienza University and INFN for the support with funding and the CMS MTD collaboration for discussion and feedback on the measurements.









\bibliographystyle{JHEP}
\bibliography{mybib}

\end{document}